\documentclass[lettersize,journal]{IEEEtran}
\usepackage{cite}
\usepackage{amsmath,amssymb,amsfonts}
\usepackage{algorithmic}
\usepackage{textcomp}
\usepackage{wrapfig}
 
\usepackage{multirow}
\usepackage{siunitx}
\usepackage{tikz}
 
\usepackage{makecell}
\usepackage{xurl}
 
\usepackage{graphicx}
\usepackage[table]{xcolor}
\usepackage{colortbl}
 
\usepackage{caption}
\usepackage{subcaption}
\usepackage{comment}
\usepackage{tabularx}
 
\usepackage{adjustbox}
\usepackage{array}
\usepackage{booktabs} %for top, middle and bottomline
\usepackage{nicematrix}
\usepackage{xfrac}
\newcommand{\rr}{\ensuremath{\boldsymbol{r}}}

\begin{document}

%\linenumbers

\title{Improved Monitoring of Honey bee Colony Strength via Audio IoT Sensors, Modulation Tensorgrams and Recurrent Neural Networks}
\author{Mahsa Abdollahi, Yi Zhu, Heitor R. Guimarães, Nico Coallier, Ségolène Maucourt, Pierre Giovenazzo, and Tiago H. Falk %\IEEEmembership{Member, IEEE}
\thanks{The authors acknowledge funding from NSERC via their Alliance program (ALLRP 548872-19), as well as Nectar Technologies Inc for the support with data collection. }
\thanks{M.A., Y. Z., H. G., and T. F. are with the INRS-EMT, Université du Québec, Montréal, Canada. (e-mail: mahsa.abdollahi@inrs.ca, Yi.Zhu@inrs.ca, Heitor.Guimaraes@inrs.ca, Tiago.Falk@inrs.ca). }
\thanks{S. M., P. G. are with the biology department, Laval university, Quebec, Canada. (e-mail: segolene.maucourt.1@ulaval.ca, pierre.giovenazzo@bio.ulaval.ca).}
\thanks{N. C. is with the Nectar Technologies Inc., Montréal, Canada. (e-mail: nico@nectar.buzz).}}

%\markboth{Journal of \LaTeX\ Class Files,~Vol.~18, No.~9, September~2020}%
%{How to Use the IEEEtran \LaTeX \ Templates}

\maketitle

\begin{abstract}
Honey bees ({\it{Apis mellifera}}) play a crucial role in agriculture and ecosystem stability as key pollinators of crops and wild plants. As such, monitoring hive strength remotely with Internet of Things (IoT) sensors has become a crucial task. Previously, handcrafted features extracted from the modulation spectrum of audio IoT devices were shown to improve acoustic monitoring of colony strength. In this paper, we hypothesize that important discriminative information is present in the temporal dynamics of the modulation spectrum, but this information is discarded with prior methods. As such, we explore the use of a new modulation tensorgram where the time dimension is kept. This new representation is used as input to a convolutional neural network (CNN) and a convolutional recurrent deep neural networks (CRDNN). Using the public UrBAN dataset, which contains more than 3,000 hours of beehive audio recordings, we show that the proposed method improves both accuracy and cross‑hive generalizability over prior benchmark methods, and the results further suggest improved robustness to noisy in‑the‑wild recording conditions. We use saliency maps and gradient-weighted class activation maps for explainability and show the importance of the modulation spectral temporal dynamics for the task at hand. Overall, our results suggest that accurate, generalizable, and robust acoustic monitoring of honey bee colony strength is possible. 

\end{abstract}


\section{Introduction}
\label{introduction}

\begin{figure*}
\centering
\includegraphics[width=0.7\linewidth]{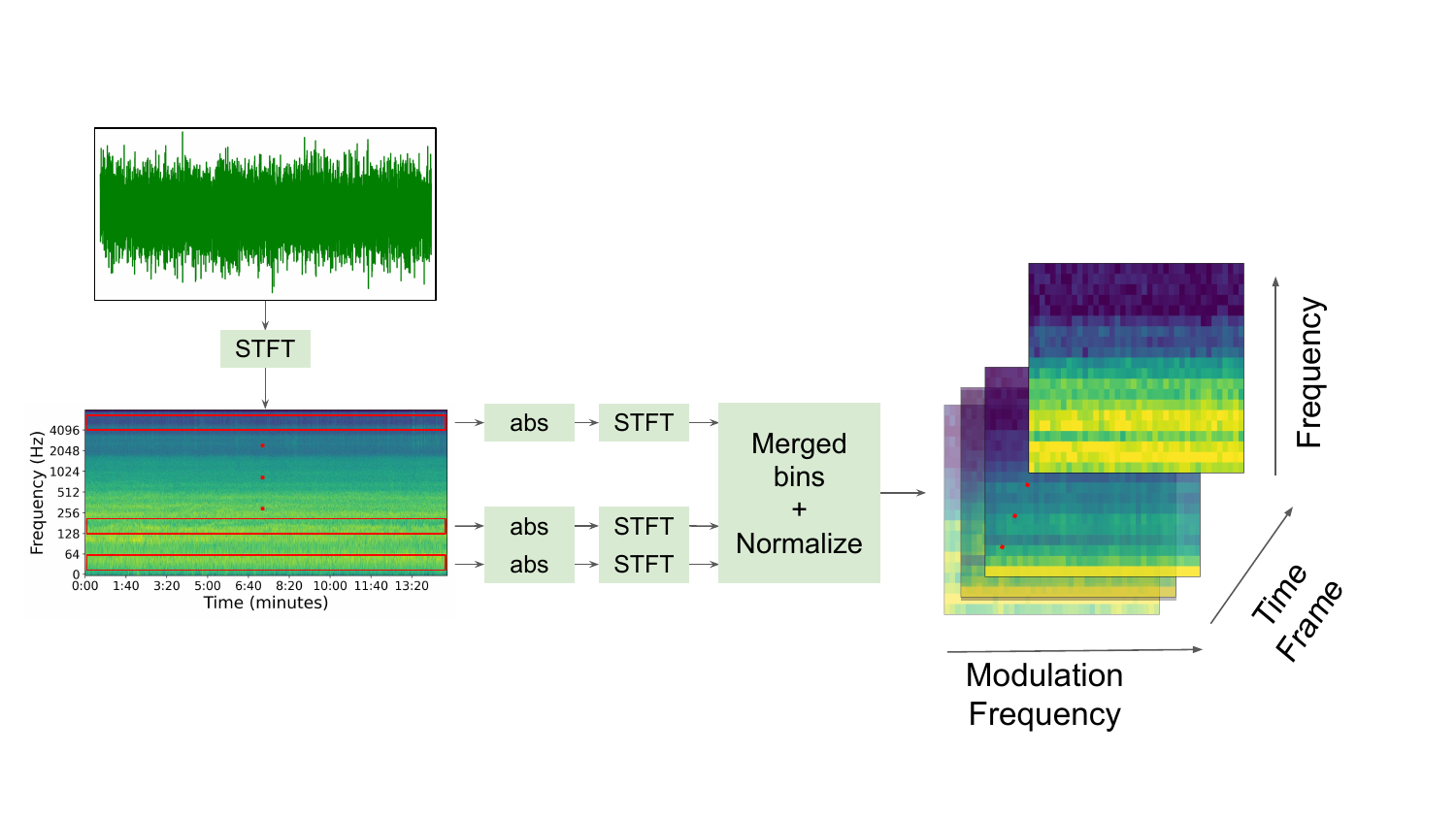}
\caption{Block diagram of signal processing steps involved in calculation of the modulation tensorgram.} \vspace{-4mm}
\label{fig:modulation_diagram}
\end{figure*}

Honey bees (Apis mellifera) play a vital role as pollinators in both agricultural and natural ecosystems, supporting the reproduction of crops and wild plants alike~\cite{potts2016safeguarding}. Despite their ecological and economic significance, global honey bee populations are declining at an alarming rate due to a combination of stressors, including pesticide use, parasitic mites, disease, and climate change~\cite{singh2025honeybees, mahankuda2024impact}. These losses pose a serious threat to biodiversity and food security~\cite{vanbergen2013threats}. Traditional hive assessments typically rely on manual inspections conducted at multi-week intervals, an approach that is not only time-intensive and laborious but can also be disruptive to the colony. Moreover, for commercial beekeepers managing hundreds or thousands of colonies, grading colony strength at scale using manual methods is practically infeasible. This creates significant management challenges because critical decisions, including mite treatment, supplemental feeding, and the selection of colonies for pollination contracts, depend heavily on accurate and timely assessments of colony strength. These constraints highlight the need for non-invasive and continuous monitoring systems capable of providing real-time and objective insights into hive health.

Advances in precision apiculture, particularly through the integration of Internet-of-Things (IoT) technologies, have enabled the development of automated systems for hive monitoring~\cite{zacepins2012application, alleri2023recent}. These sensor-based platforms capture internal hive parameters such as temperature, humidity, and acoustic activity, transmitting data remotely to facilitate timely detection of critical events like pest infestation, queen loss, or behavioral abnormalities~\cite{chen2024machine, barbisan2024machine}. Among these modalities, acoustic monitoring has emerged as especially valuable. Honey bees generate a rich variety of sounds via wingbeats, thoracic vibrations, and other movements, offering a non-invasive window into the colony health state~\cite{hunt2013intracolony}. In-hive microphones can record such buzzing sounds, which have been used to detect the presence of the queen bee~\cite{robles2024convolutional, rustam2024bee} and parasite infestations~\cite{chen2024machine, sc_34, 11409881}, detect swarming~\cite{janetzky2023swarming}, predict winter survivability~\cite{zhu2023bee}, and measure variations in colony strength~\cite{zhu2024mspb, abdollahi2022importance}.

Beyond event detection, continuous acoustic monitoring also enables the observation of behavioral shifts within the colony. Because honey bees communicate primarily through pheromones and vibrational signals, stressors such as hornet attacks or Varroa mite infestations can significantly alter the acoustic profile of the hive~\cite{mattila2021giant, 11409881}. For example, defensive or alarm-related behaviors, including piping signals, may shift the distribution of frequencies toward higher ranges under predator or parasite pressure. Long-term audio monitoring further facilitates the characterization of circadian rhythms and daily activity patterns, providing deeper insights into colony dynamics and overall physiological state~\cite{cejrowski2020buzz}.

The performance of such systems depends heavily on the quality of the recorded audio data. Data quality has a direct impact on the features to be extracted and used for hive monitoring. Commonly, time-frequency based methods have been proposed and the developed monitoring tools have greatly relied on spectrogram~\cite{ruvinga2022prediction, sc_20} and mel-frequency cepstral coefficient (MFCC)\cite{rustam2024bee, phan2023investigation} based features. MFCCs, for example, have been widely used in bee species recognition, swarming, and queen presence detection\cite{abdollahi2022automated}. These time-frequency based audio methods, however, are very sensitive to overlapping noise sources, such as beekeeper speech, rain, wind, and other nearby factors, such as trains, tractors\cite{abdollahi2022importance}.

Moreover, while traditional hive monitoring tools relied on conventional machine learning classifiers \cite{abdollahi2022automated}, recently, there has been a rise on the use of deep neural networks, specifically convolutional neural networks (CNNs). CNNs were originally developed for image-like inputs, thus are well suited for spectrograms or mel-scaled spectrograms. The work in \cite{ieee_16} employed a CNN with four convolutional layers to detect queenless hives using the spectrogram as input. In another study, CNNs were successfully applied to detect honey bee piping signals, a key indicator of colony reproductive dynamics~\cite{campell2025honey}. Similarly, the work in \cite{sc_3} showed that CNN-based classifiers outperformed traditional machine learning models across several beehive monitoring tasks.

In parallel to these developments, the so-called modulation spectral features emerged to better capture the rhythmic and temporal patterns in beehive acoustics. These features highlight modulations in signal energy across time and frequency, making them well-suited for identifying the repetitive vibratory patterns typical of bee colonies and separating these patterns from unwanted noise sources and artifacts. In prior work, we proposed to handcraft some features from the modulation spectrum and we showed that such features were complementary to spectrogram-based ones using traditional machine learning classifiers, such as random forests~\cite{abdollahi2025audio}. Here, we build on this foundational work and explore the use of the raw modulation spectrogram as input to a CNNs and a convolutional recurrent deep neural networks (CRDNNs). This approach allows us to exploit the spatial and temporal richness of the modulation representations, as well as their noise-robustness properties, while leveraging the modeling power of deep neural networks. Our experiments show the proposed method generalizing to unseen conditions more efficiently than several benchmark methods, thus helping address a critical issue in remote beehive monitoring - generalizability across unseen environments~\cite{bricout2024bee, ieee_16}. 

The novelty of the present study lies in three main aspects. First, we move beyond handcrafted modulation descriptors by directly using raw modulation spectrograms and time-preserving modulation tensorgrams as structured inputs to deep learning models, thereby preserving richer temporal–spectral information that is often lost in aggregated feature representations.  Second, we provide a systematic evaluation of multiple deep learning architectures, including 2D CNNs, 3D CNNs, and CRDNNs, for the task of colony strength regression, allowing a comprehensive assessment of how different model capacities and inductive biases interact with modulation-based representations. Third, we incorporate explainability analyses using saliency maps and gradient-weighted class activation maps (Grad-CAM) visualizations to interpret the learned representations and to identify which time–frequency–modulation regions contribute most to the predictions, improving the transparency and interpretability of the proposed approach.

The remainder of this paper is organized as follows. Section~\ref{method} describes the proposed method. Section~\ref{experiment} describes the experimental setup, while Section~\ref{results} discusses the obtained results. Lastly, Section~\ref{conclusion} presents the conclusions.

\section{Proposed Method}\label{method}
\subsection{Modulation Spectrum}

In many acoustic classification tasks, traditional spectrograms struggle in noisy environments. This limitation arises because both signal and noise are distributed across the same time-frequency space, making separation difficult. To address this limitation, the modulation spectrogram and its time-resolved counterpart, the modulation tensorgram, have been proposed~\cite{tiwari2022modulation}. Figure~\ref{fig:modulation_diagram} depicts the signal processing steps involved in the computation of the modulation tensorgram.

%First, a short-time Fourier transform (STFT) with a 100 ms window length and a 12.5 ms hop length is used, resulting in the widely-used spectrogram. Next, for each frequency bin, the spectral magnitude is taken, followed by another STFT across the time dimension, with a 60 s window length and no overlap operating over the spectral magnitude time series. These window and hop lengths were optimized empirically based on pilot experimentation. The end result is a 3-dimensional tensor where the y-axis corresponds to the conventional acoustic frequency, the x-axis to the so-called modulation frequency, which measures the rate of change of each spectral bin, and the z-axis to time, where each $freq \times modulation frequency$ modulation spectrogram is computed over the second STFT frame. This 3-dimensional $freq \times modulation frequency \times time$ representation is referred to as a modulation tensorgram. If we average over the time dimension, we are left with the 2-dimensional $freq \times modulation frequency$ representation, referred to as a modulation spectrogram. 

First, a short-time Fourier transform (STFT) with a 100 ms window length and a 12.5 ms hop length is applied, resulting in the conventional spectrogram. Next, for each acoustic frequency bin independently, the spectral magnitude time series is extracted and a second STFT is computed across the time dimension using a 60 s window with no overlap, resulting in an intermediate tensor of $(freq \times modulation\ frequency \times time) = (801 \times 801 \times 14)$. These window and hop lengths were optimized empirically based on pilot experimentation.  For the short-term STFT, window lengths of 50 ms, 100 ms, and 200 ms were evaluated. Smaller windows (50 ms) provided higher temporal resolution but poorer frequency resolution, while larger windows (200 ms) improved frequency resolution at the expense of temporal precision and smoothed short acoustic fluctuations. A 100 ms window provided the best trade-off for capturing hive acoustic activity. Different hop sizes were also tested, since smaller hop lengths improve temporal continuity and allow higher observable modulation frequencies, whereas larger hop sizes reduce modulation resolution. A hop size of 12.5 ms yielded the most stable modulation representations.

For the long-term modulation analysis, aggregation windows ranging from 30 s to 5 min were explored. Shorter windows were more sensitive to transient environmental events, while larger windows excessively smoothed colony acoustic dynamics. A 60 s aggregation window provided the best balance between robustness and temporal specificity. Finally, no overlap was applied at the long-term aggregation stage to reduce computational redundancy and simplify downstream processing.

The resulting representation is a 3-dimensional tensor where the y-axis corresponds to acoustic frequency, the x-axis to modulation frequency (i.e., the rate of temporal variation within each acoustic frequency bin), and the z-axis to time, where each $freq \times modulation\ frequency$ slice corresponds to one second-stage STFT frame. This $freq \times modulation\ frequency \times time$ representation is referred to as a modulation tensorgram. Averaging across the time dimension yields the 2-dimensional modulation spectrogram representation.

Prior to bin merging, the acoustic-frequency axis was capped at 1000 Hz. This choice is motivated by honey bee bioacoustics: the wingbeat, thoracic, and buzzing sounds that characterize in-hive activity concentrate at low frequencies, with their fundamental and lower harmonic components falling well below 1000 Hz. The discriminative analysis presented later in Section~\ref{disc_ms} is consistent with this choice, with the most discriminative regions appearing in the lower portion of this range. To further reduce dimensionality, the retained acoustic-frequency bins were then grouped down to 20, and adjacent modulation-frequency bins were grouped to 40, resulting in a final tensor size of $20 \times 40 \times 14$ for the modulation tensorgram and $20 \times 40$ for the modulation spectrogram.

\begin{figure*}%[H]
\centering
\subfloat[]{\label{fig:}
\centering
\includegraphics[width=0.3\linewidth]{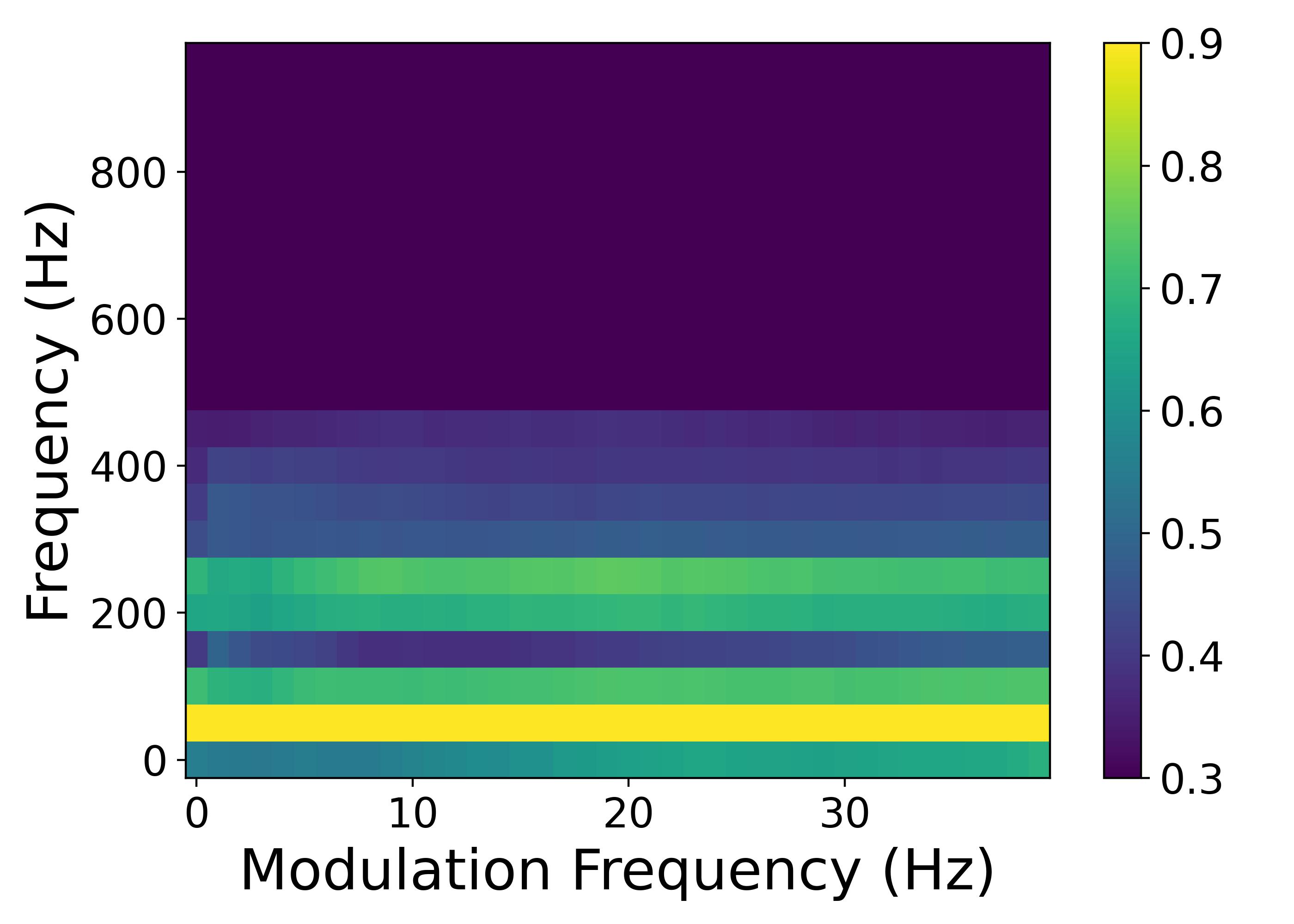}
}
\subfloat[]{\label{fig:}
\centering
\includegraphics[width=0.3\linewidth]{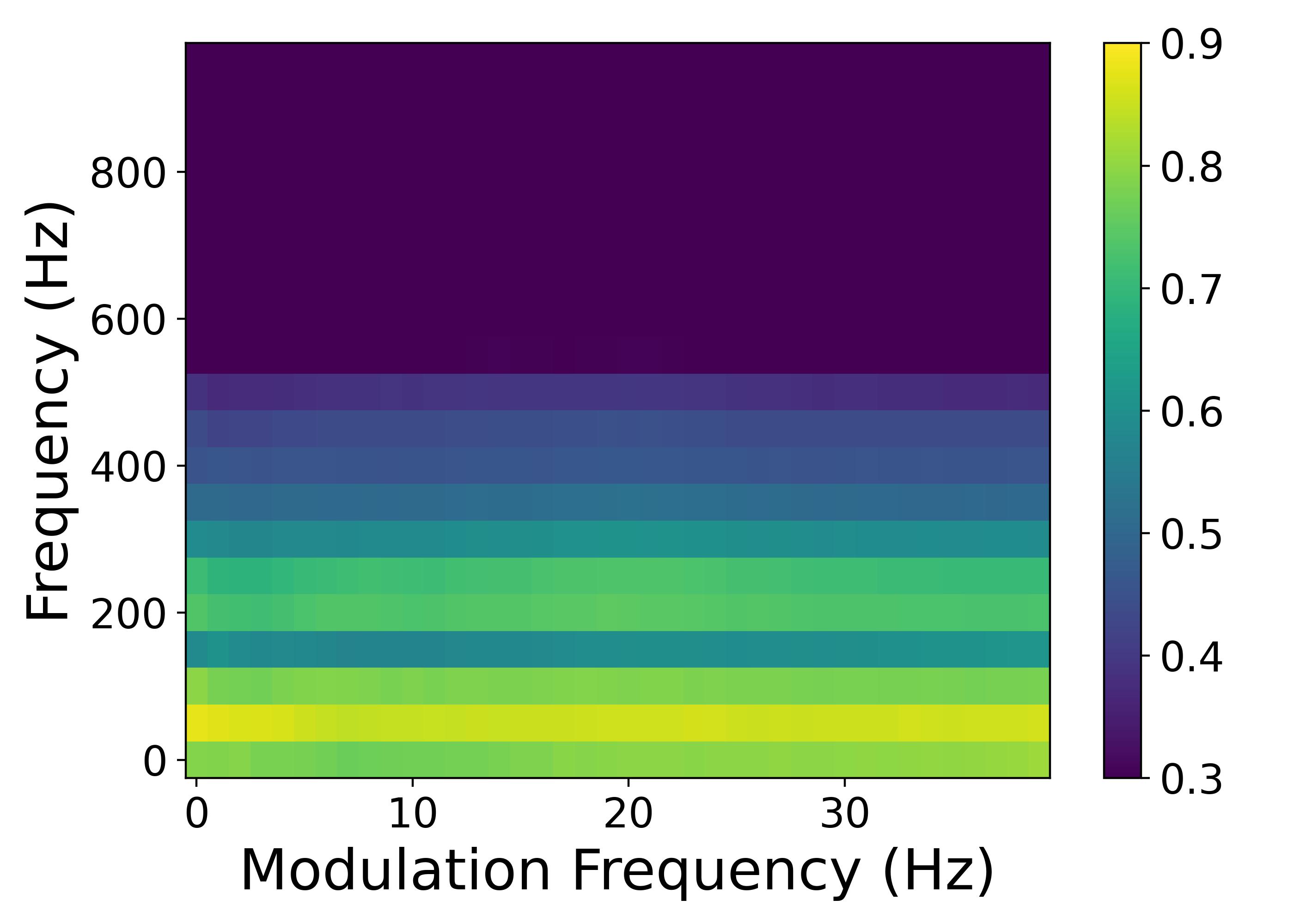}
}
\subfloat[]{\label{fig:}
\centering
\includegraphics[width=0.3\linewidth]{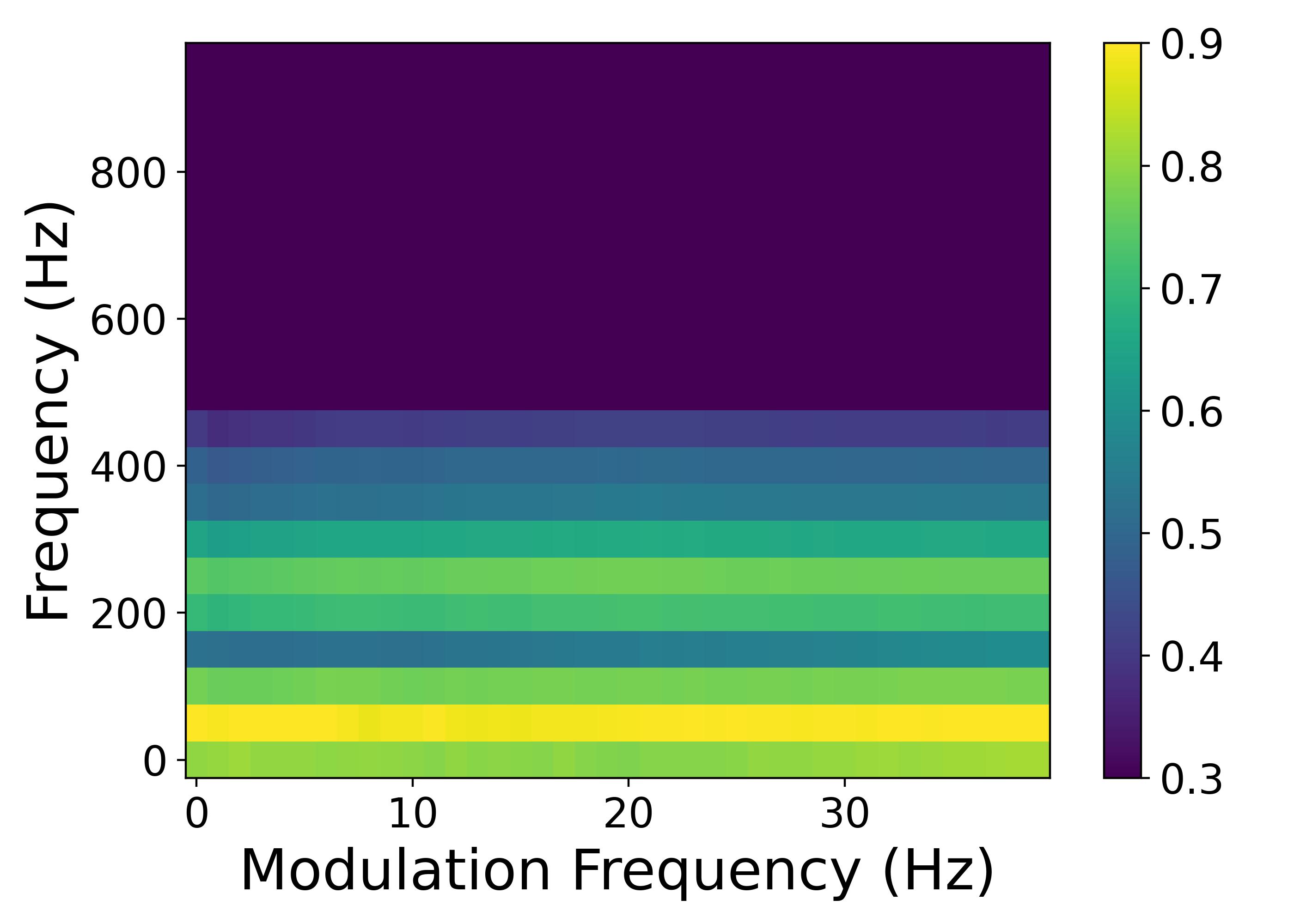}
}
\caption{Modulation spectrograms averaged on samples with a) $1 \leq \textit{fobs} \leq 10$, b) $11 \leq \textit{fobs} \leq 20$, and c) $21 \leq \textit{fobs} \leq 30$. }
\label{fig:mod_2d_avg}
\end{figure*}

To illustrate some representative modulation spectrograms, Figure~\ref{fig:mod_2d_avg} (a)-(c) shows an average representation for beehives of varying strengths, namely with frames of bees ($fob$) ranging from $1 \leq \textit{fobs} \leq 10$, $11 \leq \textit{fobs} \leq 20$, and $21 \leq \textit{fobs} \leq 30$, respectively. %For these plots, all audio recordings were taken between 8-11pm. 
For visualization purposes, these examples were selected from recordings collected between 8 pm and 11 pm, a period corresponding to peak hive occupancy when the majority of forager bees have returned to the colony. This time window provides acoustic patterns that are more representative of the collective colony state and less influenced by transient variability associated with daytime foraging activity.

\subsection{Discriminative power of the modulation spectrum}\label{disc_ms}

To measure the discriminative power of the modulation spectrogram, we compute the Fisher ratio (F-ratio) between different hive strength groups. The F-ratio quantifies how well a given feature distinguishes between two classes by comparing two types of variance: the between-group variance, which captures the separation between group means, and the within-group variance, which reflects variability within each group. 
%The F-ratio is defined as:
%\begin{equation}
%    F = \frac{\sigma_{\text{between}}^2}{\sigma_{\text{within}}^2},
%\end{equation}
%where ${\sigma_{\text{between}}^2}$ is the variance of the feature means between the groups, and ${\sigma_{\text{within}}^2}$ is the average variance within each group. 
A higher F-ratio value indicates greater discriminability. 

\begin{figure*}%[H]
\centering
\subfloat[]{\label{fig:}
\centering
\includegraphics[width=0.3\linewidth]{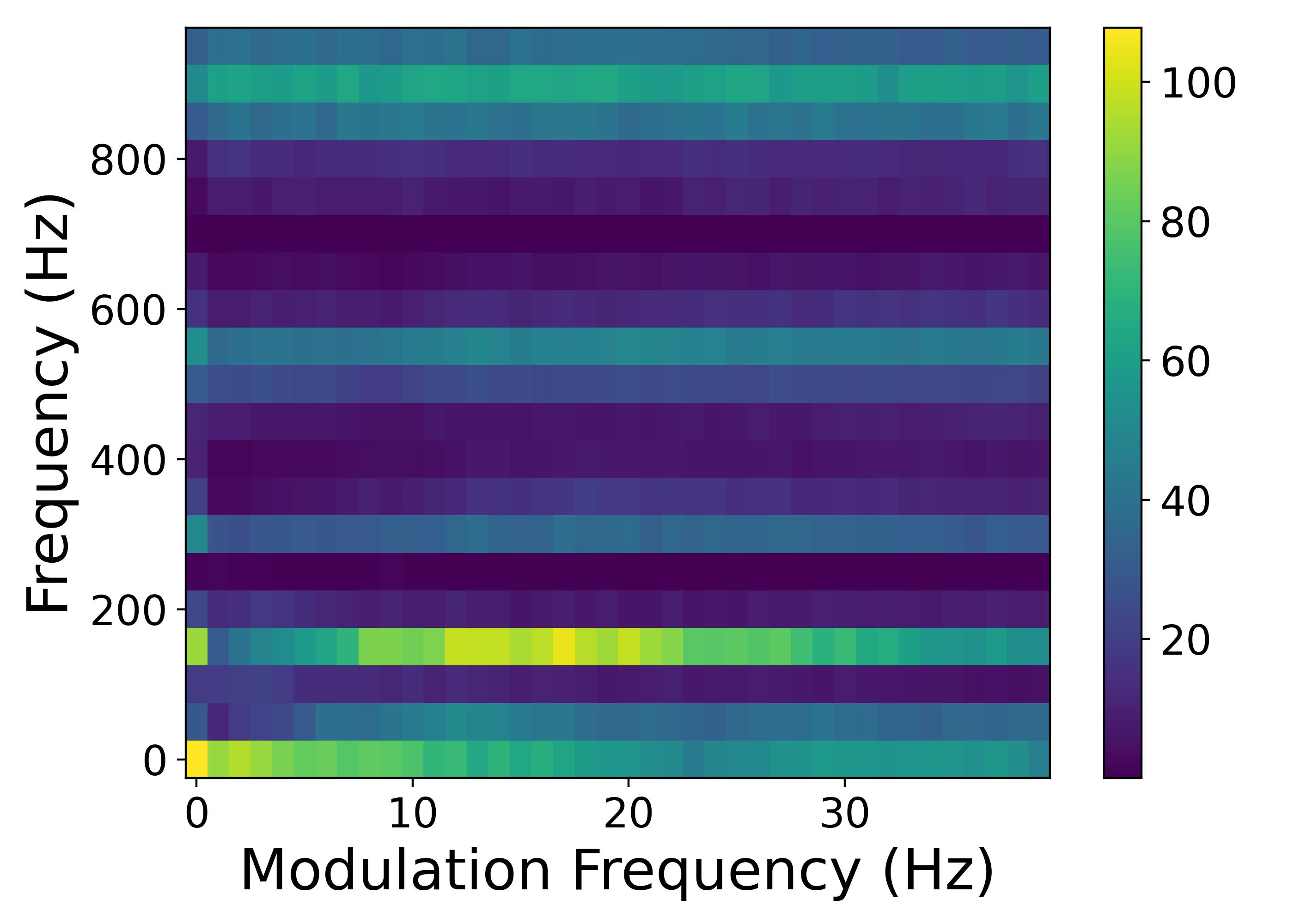 }
}
\subfloat[]{\label{fig:}
\centering
\includegraphics[width=0.3\linewidth]{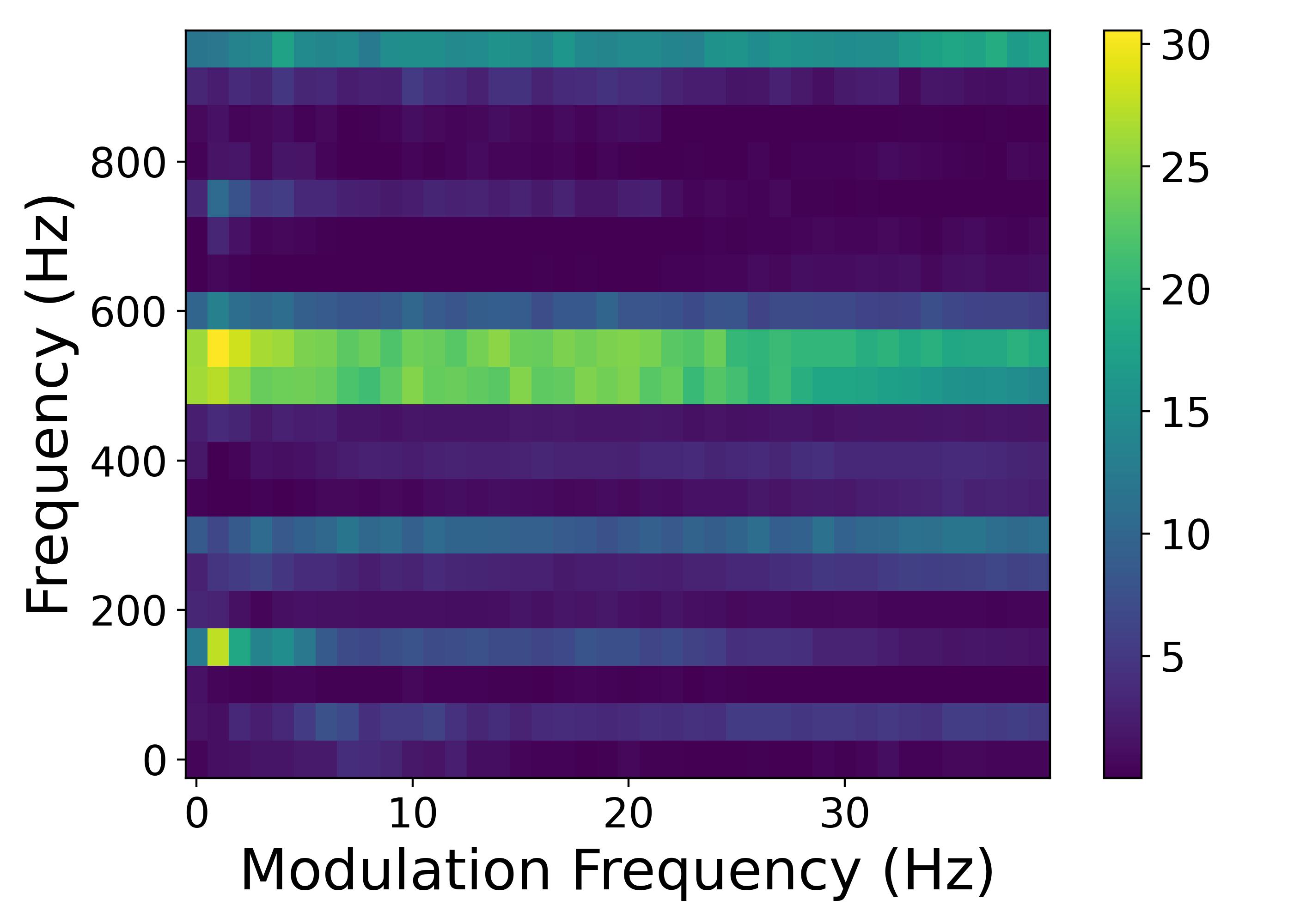}
}
\subfloat[]{\label{fig:}
\centering
\includegraphics[width=0.3\linewidth]{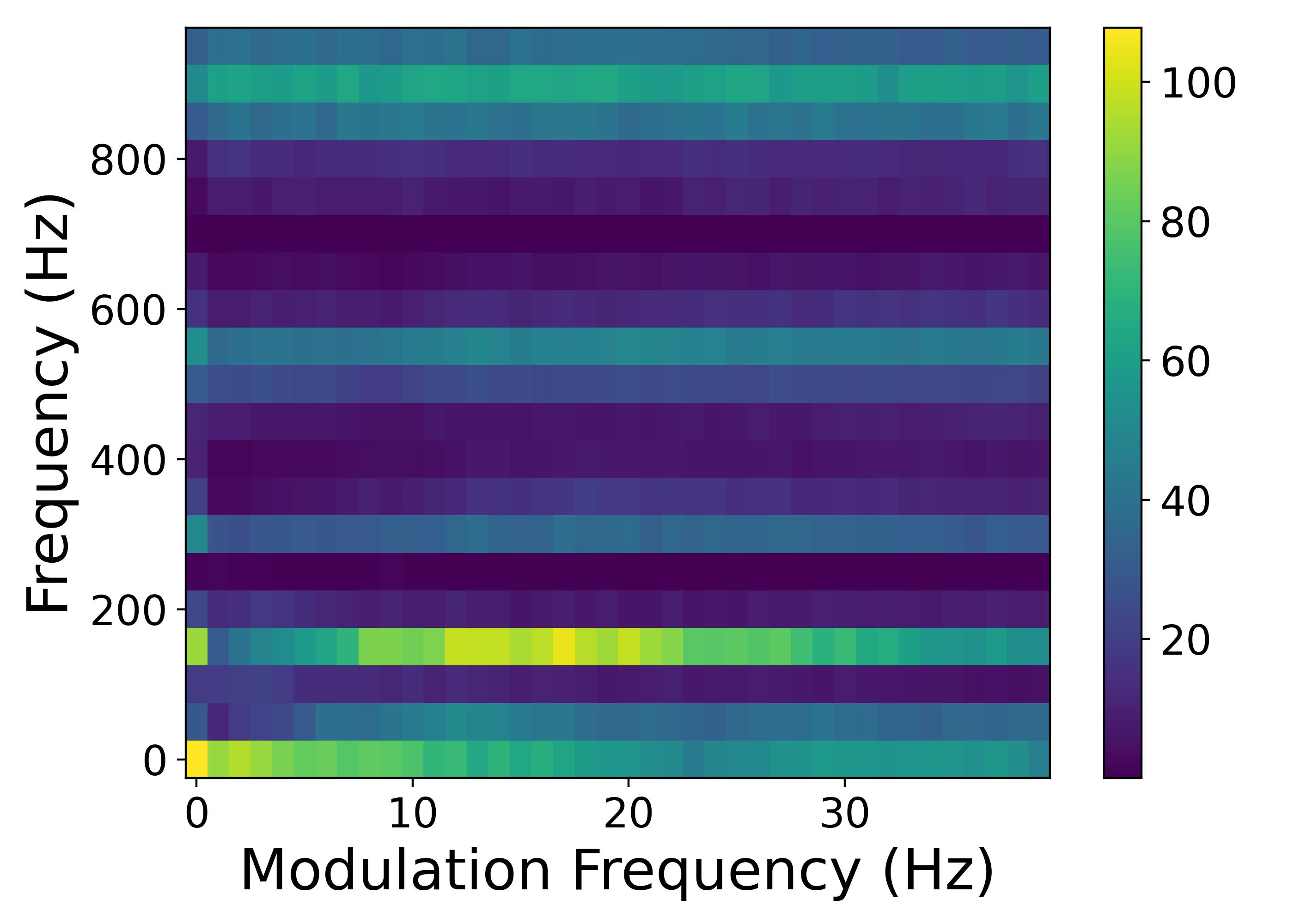}
}
\caption{The number of frames of bees (\textit{fobs}) is grouped into three categories: Category 1 corresponds to $1 \leq \textit{fobs} \leq 10$, Category 2 to $11 \leq \textit{fobs} \leq 20$, and Category 3 to $21 \leq \textit{fobs} \leq 30$. F-ratio maps comparing the following pairs of categories: a) Category 1 vs. Category 2, b) Category 2 vs. Category 3, and c) Category 1 vs. Category 3. }
\label{fig:f_ratio} \vspace{-4mm}
\end{figure*}

To compute the F-ratio, the number of frames of bees (\textit{fobs}) was divided into three population categories: Category~1 includes \(1 \leq \textit{fobs} \leq 10\), Category~2 includes \(11 \leq \textit{fobs} \leq 20\), and Category~3 includes \(21 \leq \textit{fobs} \leq 30\). These three categories correspond to the conditions shown in Figures~\ref{fig:mod_2d_avg} (a)-(c). Figure~\ref{fig:f_ratio} presents the resulting F-ratio maps for the following pairwise comparisons: (a) Category~1 vs.~Category~2, (b) Category~2 vs.~Category~3, and (c) Category~1 vs.~Category~3. These maps highlight the modulation–spectral regions where the differences in energy are most pronounced across varying levels of bee activity. In particular, the lower-frequency regions around 150--200~Hz, as well as the band near 500--600~Hz, show the strongest discriminative power, indicating that changes in these modulation components should be closely associated with increases in colony activity and could be leveraged for automated beehive strength monitoring tasks.

\subsection{Deep neural network model architectures}
To model the acoustic patterns of beehive audio, we implemented several deep learning architectures capable of learning from 2D and 3D input representations. These models are optimized to handle spectrograms, modulation spectrograms, and modulation tensorgrams. The architectures tested include a baseline CNN, a CNN with spatial attention, a convolutional-recurrent deep neural network (CRDNN), and a 3D variant for modulation tensorgrams. 
Models were trained using the Keras framework~\cite{gulli2017deep} for 300 epochs using the Adam optimizer. All models were trained using the RMSE as the loss function. An initial learning rate of 0.0001 was employed, together with an adaptive learning rate strategy to improve convergence during training. Experiments were conducted with batch sizes of 128.
Early stopping was used to monitor the validation loss and terminate training if no meaningful improvement (minimum change of $1 \times 10^{-4}$) was observed over 15 consecutive epochs. The best-performing model weights were restored at the end of training.
In addition, a ReduceLROnPlateau strategy was applied to adaptively decrease the learning rate by a factor of 0.5 when the validation loss plateaued for 5 epochs, with a lower bound of $1 \times 10^{-9}$, ensuring continued but controlled learning.
More details on these architectures and benchmark systems is given next.

\subsubsection{Baseline CNN}
The baseline convolutional model consists of three convolutional layers using $3\times3$ kernels and ReLU activation. Each block is followed by batch normalization and pooling, with L1 regularization applied to the deeper layers to reduce overfitting. The final convolutional features are flattened and passed through a dense layer with 64 units, followed by dropout, and finally mapped to a scalar regression output. Details about this architecture are given in Table~\ref{tab:cnn_baseline}.

\begin{table}%[ht]
\small
\centering
\caption{Details of baseline CNN model}
\label{tab:cnn_baseline}
\begin{tabular}{p{2.2cm}p{3.8cm}}
\toprule
\textbf{Layer} & \textbf{Details} \\
\midrule
Conv2D & 32 filters, $3\times3$, ReLU, same padding \\
BatchNorm & -- \\
MaxPool2D & $2\times2$ pool \\
Conv2D & 64 filters, $3\times3$, ReLU, L1 reg, same padding \\
BatchNorm & -- \\
MaxPool2D & $2\times2$ pool \\
Conv2D & 128 filters, $3\times3$, ReLU, L1 reg, same padding \\
BatchNorm & -- \\
Dropout & Rate 0.25 \\
Flatten & -- \\
Dense & 64 units, ReLU \\
BatchNorm & -- \\
Dropout & Rate 0.5 \\
Dense & 1 unit, linear activation \\
\bottomrule
\end{tabular}%\vspace{-2mm}

\end{table}
\subsubsection{CNN with Spatial Attention}
To enhance model focus on salient acoustic regions, spatial attention modules were integrated into the CNN pipeline (see Table~\ref{tab:cnn_attention} for more details). Attention maps are computed using a $7\times7$ convolution after each convolutional layer, emphasizing relevant feature regions. This attention-guided feature enhancement improves robustness to irrelevant background noise, while L2 regularization and dropout are employed to reduce overfitting.

\begin{table}%[ht]
\small
\centering
\caption{Details of CNN model with spatial attention}
\label{tab:cnn_attention}
\begin{tabular}{p{2.2cm}p{3.8cm}}
\toprule
\textbf{Layer} & \textbf{Details} \\
\midrule
Conv2D & 32 filters, $3\times3$, ReLU, same padding \\
BatchNorm & -- \\
Spatial Attn. & 7x7 Conv2D, sigmoid \\
MaxPool2D & $2\times2$ pool \\
Conv2D & 64 filters, $3\times3$, ReLU, L2 reg \\
BatchNorm & -- \\
Spatial Attn. & 7x7 Conv2D, sigmoid \\
MaxPool2D & $2\times2$ pool \\
Conv2D & 128 filters, $3\times3$, ReLU, L2 reg \\
BatchNorm & -- \\
Spatial Attn. & 7x7 Conv2D, sigmoid \\
Dropout & Rate 0.25 \\
Flatten & -- \\
Dense & 64 units, ReLU \\
BatchNorm & -- \\
Dropout & Rate 0.5 \\
Dense & 1 unit, linear activation \\
\bottomrule
\end{tabular}
\end{table}

\subsubsection{CRDNN (Convolutional-Recurrent-DNN)}

To capture both local and long-range temporal dependencies in the acoustic data, we designed a CRDNN architecture (see Table~\ref{tab:crdnn_model}). The model begins with three convolutional layers, each followed by batch normalization and pooling. After flattening, the output is reshaped into a temporal sequence suitable for a Gated Recurrent Unit (GRU) processing. A two-layer GRU extracts sequential patterns, followed by fully connected layers for final prediction. This hybrid structure combines the strengths of CNNs for spatial and GRUs for temporal modeling.

\begin{table}%[ht]
\small
\centering
\caption{Details of CRDNN model}
\label{tab:crdnn_model}
\begin{tabular}{p{2.2cm}p{3.8cm}}
\toprule
\textbf{Layer} & \textbf{Details} \\
\midrule
Conv2D & 32 filters, $3\times3$, ReLU, same padding \\
BatchNorm & -- \\
MaxPool2D & $2\times2$ pool \\
Conv2D & 64 filters, $3\times3$, ReLU, L1 reg, same padding \\
BatchNorm & -- \\
MaxPool2D & $2\times2$ pool \\
Conv2D & 128 filters, $3\times3$, ReLU, L1 reg, same padding \\
BatchNorm & -- \\
Dropout & Rate 0.25 \\
Flatten & -- \\
Reshape & To $(\text{seq\_len}, 128)$ \\
GRU & 32 units, return seq = True \\
Dropout & Rate 0.5 \\
GRU & 64 units, return seq = False \\
Dense & 256 units, ReLU \\
BatchNorm & -- \\
Dropout & Rate 0.5 \\
Dense & 128 units, ReLU \\
BatchNorm & -- \\
Dropout & Rate 0.5 \\
Dense & 1 unit, linear activation \\
\bottomrule
\end{tabular}
\end{table}

\subsubsection{3D variants}
Lastly, to leverage the full dimensionality of the modulation tensorgram representations, we designed 3D variants of our convolutional and attention-based models. Unlike the 2D models, which operate on standard time-frequency spectrograms, the 3D models process volumetric inputs that capture time, frequency, and modulation frequency simultaneously. This three-dimensional structure enables the models to extract joint spatio-spectro-temporal features, making them well-suited for tasks involving modulation-based acoustic representations. The 3D CNN with attention integrates spatial attention mechanisms across the 3D input, enhancing relevant regions within the tensor while suppressing less informative ones. Each convolutional block includes batch normalization, 3D max-pooling, and dropout for regularization, followed by dense layers for regression output. By capturing patterns across all axes of the tensorgram, these models are able to exploit modulation structures more effectively, improving the robustness and generalizability of acoustic monitoring tasks.

\section{Experimental Setup}\label{experiment}

\subsection{Dataset Description}
To test the efficacy of our proposed method, we rely on the UrBAN dataset~\cite{abdollahi2025urban}, which includes over 3,000 hours of raw audio data collected from an urban apiary. The UrBAN dataset is comprised of audio data collected from nine beehives monitored over a two-year period at a rooftop apiary in Montréal, Québec, Canada. Each hive included one brood chamber and one-to-two honey supers, housed within 10-frame standard Langstroth boxes. 

\begin{figure}%[H]
\centering
\subfloat[]{\label{fig:fob_2021_detaied}
\centering
\includegraphics[width=0.9\linewidth]{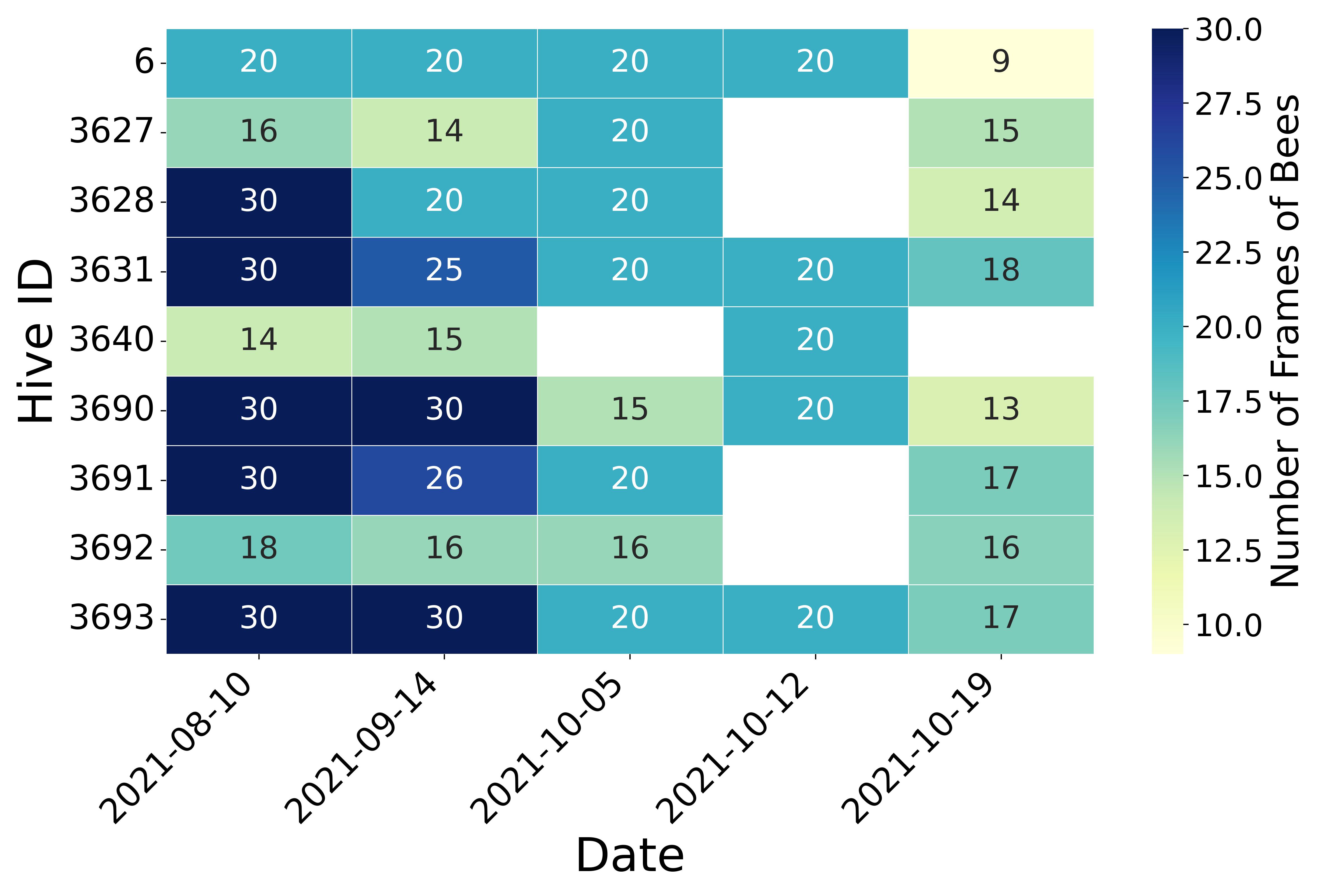}
}

\subfloat[]{\label{fig:fob_2022_detaied}
\centering
\includegraphics[width=0.9\linewidth]{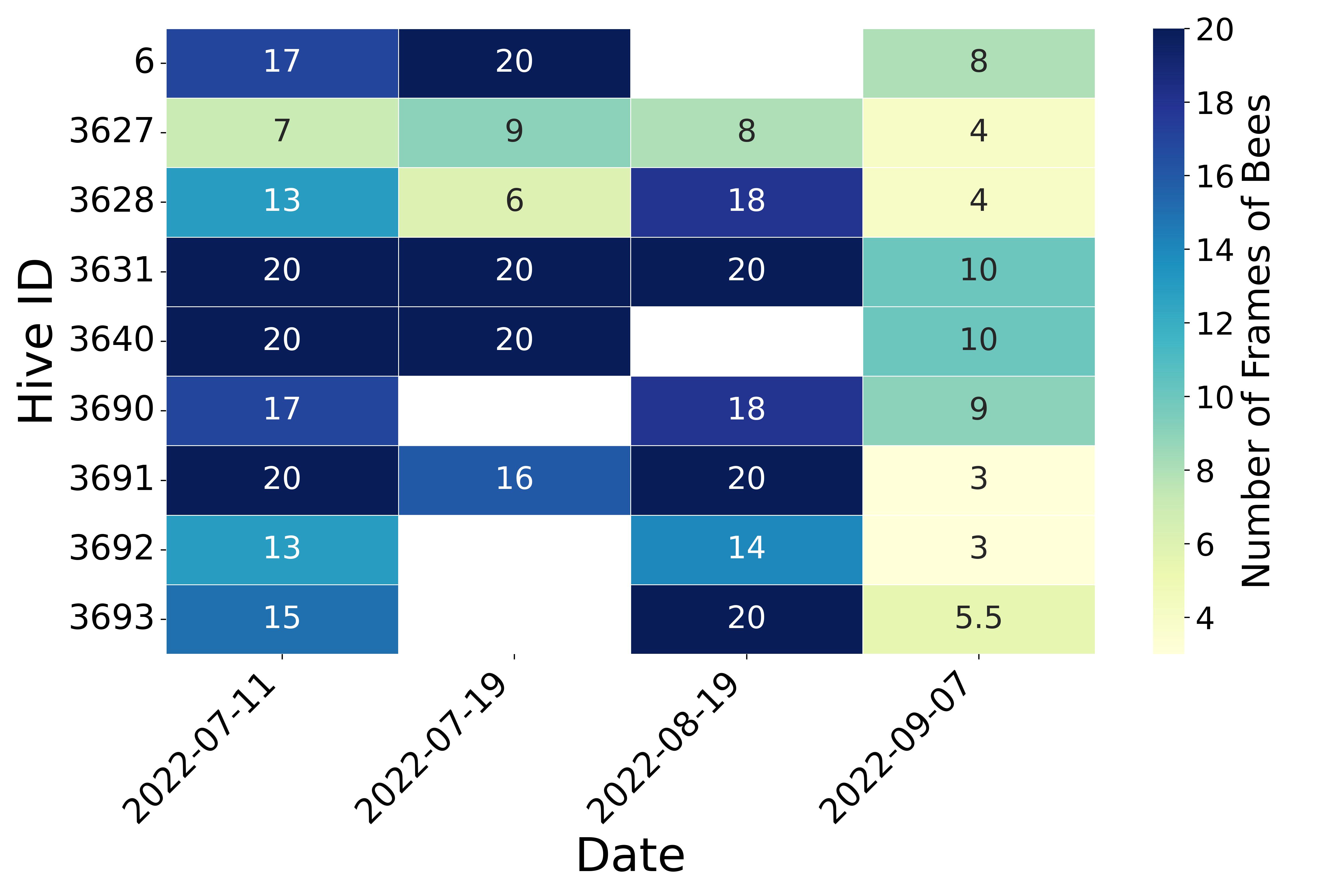}
}
\caption{Heatmaps showing the number of frames of bees estimated during visual inspection for each hive throughout (a) 2021 and (b) 2022. Values between 1-10 suggest only the brood chamber was present, 11-20 brood chamber plus one super, and 21-30 brood chamber plus two supers.}
\label{fig:fob_detailed}
\end{figure}

The hives were manually inspected to assess hive strength, verify the presence of a laying queen, and record any additional colony activity observations. Hive strength is defined as the number of frames where at least 70\% of the frame is covered by bees \cite{chabert2021rapid}. A microphone was installed atop the central frame of the bottom brood box. The audio data consists of 15-minute audio segments recorded every 30 minutes at a sampling rate of \SI{16}{\kilo\hertz}. Figure~\ref{fig:fob_detailed} shows the number of frames of bees (fob) for each beehive in 2021 (top plot) and 2022 (bottom). Values between 1-10 suggest only the brood chamber was present, while 11-20 the brood chamber plus one super, and 21-30 the brood chamber plus two supers.

Because manual inspections occur at discrete dates while audio is recorded continuously, a colony strength label was assigned to each 15‑minute audio segment by linear temporal interpolation between consecutive inspection measurements rather than by nearest‑neighbour assignment. This procedure yields temporally continuous targets that better reflect the gradual population dynamics between inspections.

\subsection{Benchmark features and noise removal}
To establish a baseline for evaluating the performance of the proposed features, we incorporated two widely recognized acoustic benchmarks into our analysis.

\textbf{MFCCs}: These features are designed to simulate the non-linear human auditory system by applying a mel-scale frequency mapping, making them highly effective for bioacoustic tasks. MFCCs are a standard benchmark in precision beekeeping and have proven valuable in characterizing bee sounds \cite{abdollahi2022automated}. For this study, we extracted 13 coefficients (12 standard coefficients plus the zeroth coefficient, which represents the signal's energy) using 26 mel-filter banks. The resulting MFCC representation has dimensions of 13 × 40, where the y-axis corresponds to the 13 cepstral coefficients and the x-axis to 40 time frames, matching the time-axis resolution of the modulation representations.

\textbf{Spectrograms}: As a fundamental representation of audio signals, spectrograms visualize the evolution of spectral content over time (time-frequency analysis). They are extensively used in beehive acoustics, notably for critical tasks such as queen bee detection \cite{robles2024convolutional} and swarming prediction \cite{ruvinga2022prediction}, making them an essential comparison point for any new monitoring method. Consistent with the modulation representations, the spectrogram frequency axis was capped at 1000 Hz, and the resulting frequency bins were grouped down to 20 bins. This yields a final spectrogram of dimension $20 \times 40$ ($frequency \times time$), matching the shape of the modulation spectrogram.

Notwithstanding, as mentioned previously, these features have been shown to be sensitive to environmental noise. As the dataset used herein was collected ``in the wild,'' it is crucial that some noise suppression be performed to ensure reliable prediction from the benchmark systems. Here, we employ a classical spectral amplitude subtraction method for noise reduction, as described in~\cite{abdollahi2025urban}.

\subsection{Testing Setup and Figures-of-Merit}
Building on recommendations from previous studies, we explored two distinct experimental configurations, namely ``random-split" and ``hive-independent".  

In the random-split approach, partitioning was performed at the level of individual 15-minute audio segments, where samples were randomly assigned to the training, validation, and testing folds while preserving the overall label distribution. A nested 5-fold cross-validation strategy was then employed to ensure robust evaluation of the models. This procedure ensured that every data point was included in the test set exactly once, reducing the risk of overfitting and providing a more reliable estimate of model performance. However, because temporally adjacent recordings may exhibit strong acoustic correlations, this setup may produce optimistic performance estimates due to similarities between training and testing samples collected from nearby time periods.

Conversely, in the hive-independent setup, the folds were constructed at the hive level such that all recordings originating from a given hive were assigned exclusively to either the training, validation, or testing subsets. This prevented any acoustic information from the same colony from appearing across splits. The 5 folds were therefore selected hive-independently for both validation and testing. This latter setup is more stringent and more closely resembles real-world scenarios where models must generalize to unseen hives and environmental conditions. As such, it provides a more realistic assessment of model robustness and generalizability to unseen environments.

As our task is a regression aimed at predicting the strength of a hive (given by its $fob$ value), two conventional figures-of-merit are utilized: mean absolute error (MAE) and Pearson correlation (\rr) between the real and predicted $fob$ values. An ideal/perfect regressor will have MAE close to zero and a correlation close to unity.

\section{Results and Discussion}\label{results}

\subsection{Colony Population Strength Prediction}
Table~\ref{tab:performance_mae_corr} presents the performance achieved by the different benchmark and proposed methods. As can be seen, in the random-split scenario, while the spectrogram and MFCC inputs yield MAE values around $3.5$ and correlation $r \approx 0.76$, the modulation-based methods are capable of reducing the MAE to around $1.5$ and elevate the correlation to $r \approx 0.95$. This underscores the hypothesis that bee acoustic activity is best characterized by temporal modulations (i.e., the rate and structure of changes in the dominant frequencies) rather than the static spectral envelope or cepstral coefficients. Moreover, the traditional spectrogram and MFCC benchmarks showed substantial gains in performance once audio enhancement was applied.
The gains brought by spectral subtraction were noticeably smaller for the proposed method than for the spectrogram and MFCC baselines. This pattern is consistent with the previously reported noise‑robustness of modulation spectral representations~\cite{abdollahi2025audio}. We note, however, that the UrBAN dataset does not include per‑segment annotations of environmental disturbances (rain, wind, vehicles, etc.), so this observation suggests — rather than directly demonstrates improved robustness; a stratified evaluation against labelled noise events is left as future work (see Section~\ref{limitations}).

Moreover, some attention‑based variants do not improve over, and in a few cases trail, their non‑attention counterparts after enhancement (for example Spec‑CNN‑att‑2D vs. Spec‑CNN‑2D in the random‑split setting). We attribute this to two effects: (i) once spectral subtraction has removed much of the broadband background, the residual input contains less irrelevant content for spatial attention to suppress; and (ii) the local musical‑noise artifacts that classical spectral subtraction can introduce in low‑energy regions are high‑contrast and may be incidentally selected by the attention maps. Consistent with this interpretation, attention continues to provide a benefit on the more noise‑resilient modulation‑based features and on the more challenging hive‑independent setting. Overall, for this setting, the CRDNN-3D model utilizing the modulation tensorgram as input resulted in the best performance.

For the more strict hive-independent setting, the performance is shown to drop for all models and features. Notwithstanding, overall, the modulation spectrogram and tensorgram inputs retain their superior performance, achieving correlations greater than 0.7, while the spectrogram and MFCCs benchmarks remain around 0.50. This resilience suggests that the underlying frequency-modulation patterns are more robust and generalizable indicators of bee activity than the raw spectral content, which is likely dominated by hive-specific artifacts. 

The CRDNN-3D achieved the lowest MAE in both scenarios (1.01 and 3.31). In the random-split setting, where the standard deviations are tight (e.g., 1.01 ± 0.08), this advantage is well separated from the competing methods. In the hive-independent setting the numerical advantage falls within the range of variability, so we motivate CRDNN-3D as our recommended model not on the basis of this margin alone, but on its consistently strong average ranks across the Friedman procedure (Table~\ref{tab:avg_ranks}) and its favorable accuracy–efficiency–size tradeoff (Section~\ref{Computation}).

\begin{table*}
\centering
\makebox[\textwidth][c]{
            \resizebox{0.7\textwidth}{!}{%
\begin{NiceTabular}{c|c|ccccc}
\toprule
& & \multicolumn{4}{c}{\textbf{Random-Split}} \\
\cmidrule{3-6}
\multirow{2}{*}{Feature} & \multirow{2}{*}{Model} &
  \multicolumn{2}{c}{No pre-processing} &
  \multicolumn{2}{c}{Spectral amplitude subtraction} \\
\cmidrule{3-6}
 & & {MAE} & {\rr} & {MAE} & {\rr } \\
  \cmidrule{1-6}

\multirow{3}{*}{Spectrogram} & CNN-2D &  \makecell{3.70 $\pm$ 0.17} & \makecell{0.76 $\pm$  0.03} & \makecell{1.71 $\pm$  0.13} & \makecell{0.83 $\pm$ 0.02}\\

& CNN-att-2D  & \makecell{3.68 $\pm$ 0.08}  & \makecell{0.75 $\pm$  0.02}& \makecell{2.17 $\pm$  0.09} & \makecell{0.76 $\pm$ 0.01 }\\

& CRDNN-2D& \makecell{3.49 $\pm$ 0.24}  & \makecell{0.78 $\pm$  0.02 }& \makecell{1.77 $\pm$ 0.05 } & \makecell{0.81 $\pm$ 0.01}\\

\cmidrule{1-6}
\multirow{3}{*}{MFCCs} & CNN-2D &  \makecell{3.62 $\pm$ 0.13} & \makecell{0.76 $\pm$   0.02} & \makecell{1.85 $\pm$ 0.25} & \makecell{0.81 $\pm$ 0.03}\\
& CNN-att-2D  & \makecell{3.72 $\pm$ 0.09}  & \makecell{ 0.75 $\pm$ 0.01}& \makecell{2.15 $\pm$ 0.17 } & \makecell{0.77 $\pm$ 0.01}\\

& CRDNN-2D&  \makecell{3.37 $\pm$ 0.05} & \makecell{0.78 $\pm$ 0.03}& \makecell{1.78 $\pm$  0.10} & \makecell{0.82 $\pm$ 0.01}\\

\cmidrule{1-6}
\multirow{2}{*}{\makecell{Modulation\\ spectrogram}} & CNN-2D & \makecell{1.54 $\pm$ 0.14}  & \makecell{0.95 $\pm$  0.01}& \makecell{1.14 $\pm$ 0.06} & \makecell{0.96 $\pm$ 0.01}\\
& CNN-att-2D  & \makecell{1.52 $\pm$ 0.16}  & \makecell{0.96 $\pm$  0.03}& \makecell{1.39 $\pm$ 0.06} & \makecell{0.97 $\pm$ 0.03}\\

\cmidrule{1-6}
\multirow{3}{*}{\makecell{Modulation\\ tensorgram}} & CNN-3D & \makecell{1.61 $\pm$ 0.10}& \makecell{0.94 $\pm$ 0.02}& \makecell{1.36 $\pm$ 0.09 } & \makecell{0.96 $\pm$ 0.01}\\

& CNN-att-3D  & \makecell{1.45 $\pm$ 0.05}  & \makecell{0.96 $\pm$ 0.02}& \makecell{1.23 $\pm$  0.06} & \makecell{0.96 $\pm$ 0.03}\\
& CRDNN-3D& \makecell{1.24 $\pm$ 0.03}& \makecell{0.95 $\pm$ 0.01}& \makecell{1.01 $\pm$ 0.08} & \makecell{0.97 $\pm$ 0.01}\\
\toprule
& & \multicolumn{4}{c}{\textbf{Hive-Independent}} \\
\cmidrule{1-6}

\multirow{3}{*}{Spectrogram} & CNN-2D & \makecell{4.87 $\pm$ 1.47} & \makecell{0.47 $\pm$ 0.22}& \makecell{4.74 $\pm$  1.46} & \makecell{0.52 $\pm$ 0.21}\\

& CNN-att-2D  & \makecell{4.60 $\pm$ 1.61} & \makecell{0.46 $\pm$ 0.21}& \makecell{4.50 $\pm$ 1.59} & \makecell{0.51$\pm$ 0.20 }\\

& CRDNN-2D& \makecell{4.55 $\pm$ 1.62}  & \makecell{0.56 $\pm$ 0.20}& \makecell{4.22 $\pm$  1.16} & \makecell{0.61 $\pm$ 0.20}\\

\cmidrule{1-6}

\multirow{3}{*}{MFCCs} & CNN-2D & \makecell{4.95 $\pm$ 1.37} & \makecell{0.50 $\pm$ 0.22}& \makecell{4.65 $\pm$  1.32} & \makecell{ 0.53 $\pm$ 0.22}\\

& CNN-att-2D  & \makecell{4.99 $\pm$ 1.45} & \makecell{0.47 $\pm$ 0.24}& \makecell{4.97 $\pm$ 1.46 } & \makecell{0.50 $\pm$ 0.22}\\

& CRDNN-2D& \makecell{4.42 $\pm$ 1.68}  & \makecell{0.52 $\pm$ 0.25}& \makecell{4.38 $\pm$  1.68} & \makecell{0.54 $\pm$ 0.20}\\

\cmidrule{1-6}

\multirow{2}{*}{\makecell{Modulation\\ spectrogram}} & CNN-2D & \makecell{3.48} $\pm$ 0.90 & \makecell{0.69 $\pm$ 0.13} & \makecell{3.41 $\pm$  0.87} & \makecell{ 0.74 $\pm$ 0.11}\\

& CNN-att-2D  & \makecell{3.66 $\pm$ 1.09}  & \makecell{0.72 $\pm$ 0.09} & \makecell{3.56 $\pm$ 1.05 } & \makecell{0.76 $\pm$ 0.08}\\

\cmidrule{1-6}
\multirow{3}{*}{\makecell{Modulation\\ tensorgram}} & CNN-3D & \makecell{3.84 $\pm$ 1.21}  & \makecell{0.65 $\pm$ 0.22}& \makecell{3.69 $\pm$  1.60} & \makecell{0.69 $\pm$ 0.21 }\\

& CNN-att-3D  & \makecell{3.76 $\pm$ 1.56 }  & \makecell{0.71 $\pm$ 0.15}& \makecell{3.67 $\pm$ 1.55 } & \makecell{ 0.76 $\pm$ 0.16}\\

& CRDNN-3D& \makecell{3.42 $\pm$ 1.37} & \makecell{0.72 $\pm$ 0.20}& \makecell{3.31 $\pm$ 1.36 } & \makecell{0.78 $\pm$ 0.17}\\
\bottomrule
\end{NiceTabular}
}
}
\caption{Performance comparison of MAE and correlation (r) between different feature sets and data partitioning setups, with and without spectral enhancement.}
\label{tab:performance_mae_corr}
\end{table*}

\subsection{Non-parametric Statistical Analyses}
To statistically evaluate the differences in performance seen amongst the models, we employed the Friedman test~\cite{friedman1937use} across all four experimental scenarios (random-split and hive-independent, with and without audio enhancement) and two evaluation metrics. Next, we used the Nemenyi post-hoc test~\cite{nemenyi1963distribution} to conduct pairwise comparisons and identify which specific differences were statistically significant. The results of the Friedman test are summarized in Table \ref{tab:friedman_results} and showed statistically significant differences ($p < 0.05$) among the feature set and models performances across evaluation metrics.  The critical Chi-square value for this test (with $k=11$) was determined to be 18.30 at $\alpha=0.05$. Next, Table~\ref{tab:avg_ranks} reports the average ranks derived from the Friedman procedure. For error metrics such as MAE, a lower rank signifies superior performance, while for predictive metrics such as Correlation, a higher rank is better. Notably, the CRDNN-3D achieved particularly strong average ranks, especially when tested under the challenging hive-independent scenario.

\begin{table}
\centering

\begin{NiceTabular}{c|cccc}
\toprule
& \multicolumn{4}{c}{\textbf{Random-Split}} \\
\cmidrule{2-5}
\multirow{2}{*}{Metric} &
  \multicolumn{2}{c|}{No pre-processing} &
  \multicolumn{2}{c}{Spectral amplitude subtraction} \\
\cmidrule{2-5}
  & {Friedman $\chi^2$} & {p-value}  &  
    {Friedman $\chi^2$} & {p-value}  \\
\midrule
MAE          & 46.32  & 1.25e-06  &   47.30 & 8.31e-07      \\
\rr   & 46.98 & 9.48e-07  &  47.44  & 7.82e-07      \\
\midrule
& \multicolumn{4}{c}{\textbf{Hive-Independent}} \\
\cmidrule{2-5}
MAE          & 32.78  & 2e-04  &  35.90  & 8.75e-05      \\
\rr   & 37.71  & 4.24e-05  &  40.32  & 1.48e-05      \\
\bottomrule
\end{NiceTabular}

\caption{Friedman test results across evaluation metrics MAE and correlation(r) under random-split and hive-independent scenarios.}
\label{tab:friedman_results}
\end{table}

\begin{table}
\centering

\resizebox{\columnwidth}{!}{%
\begin{NiceTabular}{c|c|ccccc}
\toprule
& & \multicolumn{5}{c}{\textbf{Random-Split}} \\
\cmidrule{3-6}
\multirow{2}{*}{Feature} & \multirow{2}{*}{Model} &
  \multicolumn{2}{c}{No pre-processing} &
  \multicolumn{2}{c}{\makecell{Spectral amplitude \\ subtraction}} \\
\cmidrule{3-6}
 & & {MAE $\downarrow$} & {\rr  $\uparrow$}  & {MAE $\downarrow$} &  {\rr  $\uparrow$} \\
 
  \cmidrule{1-6}
\multirow{3}{*}{Spectrogram} & CNN-2D & 6.6 & 4.0 & 6.6 & 5.5  \\
& CNN-att-2D  & 10.7 &  2.8 &   10.7  & 1.3 \\
& CRDNN-2D & 7.9  & 3.6  & 7.9   & 3.7 \\

  \cmidrule{1-6}
\multirow{3}{*}{MFCCs} & CNN-2D & 8.4  & 2.0  & 8.4  & 3.9  \\
& CNN-att-2D  &  10.0  & 2.8 &   10.0 & 1.7 \\
& CRDNN-2D &  7.4  & 5.8 &   7.4  & 4.9 \\
\cmidrule{1-6}
\multirow{2}{*}{\makecell{Modulation\\ spectrogram}} & CNN-2D &  1.8 &  7.9 &   1.8 &   9.9 \\
& CNN-att-2D  & 4.8    & 9.4 &  4.8 & 8.16\\

\cmidrule{1-6}
\multirow{3}{*}{\makecell{Modulation\\ tensorgram}} & CNN-3D & 4.2  & 7.9 &  4.2 & 7.9  \\
& CNN-att-3D  & 2.8  & 9.9  & 2.8  & 9.0  \\
& CRDNN-3D & \textbf{1.4}   & \textbf{9.9} &   \textbf{1.4} & \textbf{10.1} \\

\toprule
& & \multicolumn{5}{c}{\textbf{Hive-Independent}} \\
  \cmidrule{1-6}

\multirow{3}{*}{Spectrogram} & CNN-2D & 9.3  & 2.5  & 9.1 &  3.4  \\
& CNN-att-2D  & 7.6  & 2.6  &  7.6  & 3.1  \\
& CRDNN-2D & 7.2  & 6.0 & 6.3 &  6.4  \\
  \cmidrule{1-6}

\multirow{3}{*}{MFCCs} & CNN-2D & 9.2  & 4.1 &  8.8 & 3.0  \\
& CNN-att-2D  & 9.1  & 3.0  & 10.1  & 2.4 \\
& CRDNN-2D &  6.6   & 3.8  & 7.4  & 3.2   \\
  \cmidrule{1-6}

\multirow{2}{*}{\makecell{Modulation\\ spectrogram}} & CNN-2D & 3.2  & 8.2 &   3.2  & 9.1 \\
& CNN-att-2D  &  3.8  & 8.7  & 3.3   & 8.6   \\

  \cmidrule{1-6}
\multirow{3}{*}{\makecell{Modulation\\ tensorgram}} & CNN-3D & 5.1 &  8.1 & 5.0  &  8.2   \\
& CNN-att-3D  & 2.9 & 9.7  &  3.2 &  9.4 \\
& CRDNN-3D &  \textbf{2.0} & \textbf{9.3}  & \textbf{2.0}  & \textbf{9.2} \\

\bottomrule
\end{NiceTabular}}
\caption{Comparison of average ranks across evaluation metrics under random-split and hive-independent scenarios, with and without audio enhancement. Bold values correspond to best methods.}
\label{tab:avg_ranks}
\end{table}

\begin{figure*}
\centering
\subfloat[]{\label{fig:}
\centering
\includegraphics[width=0.5\linewidth]{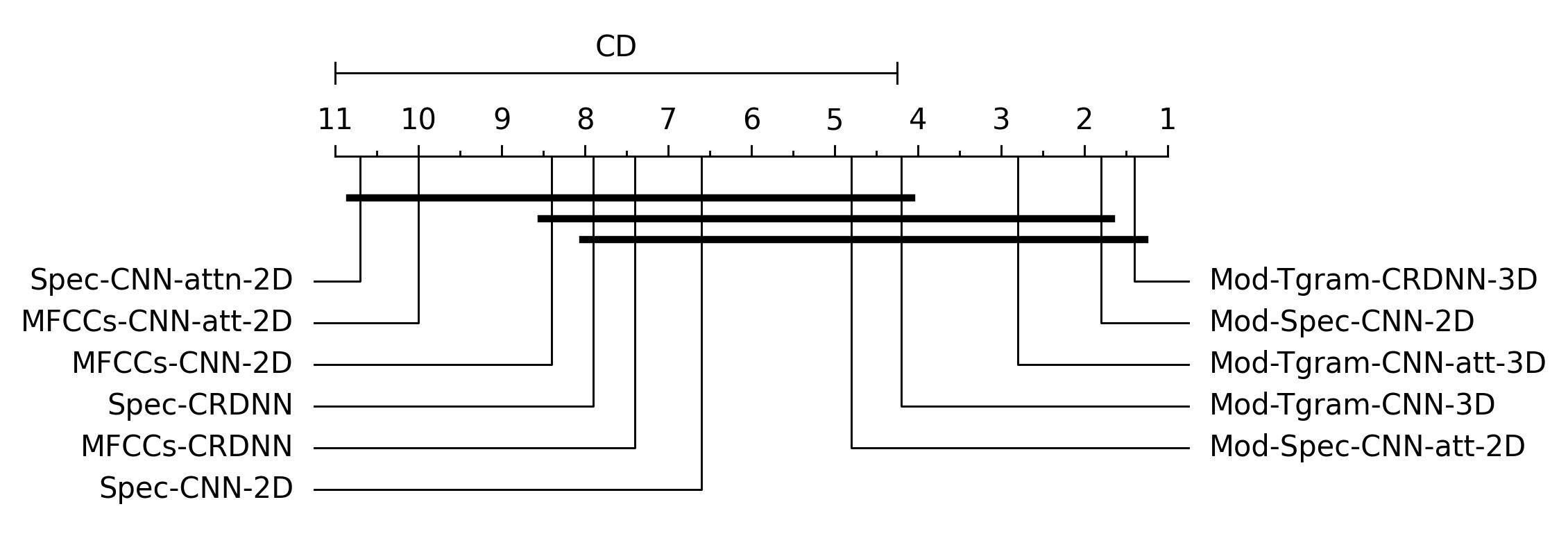}
}
\subfloat[]{\label{fig:}
\centering
\includegraphics[width=0.5\linewidth]{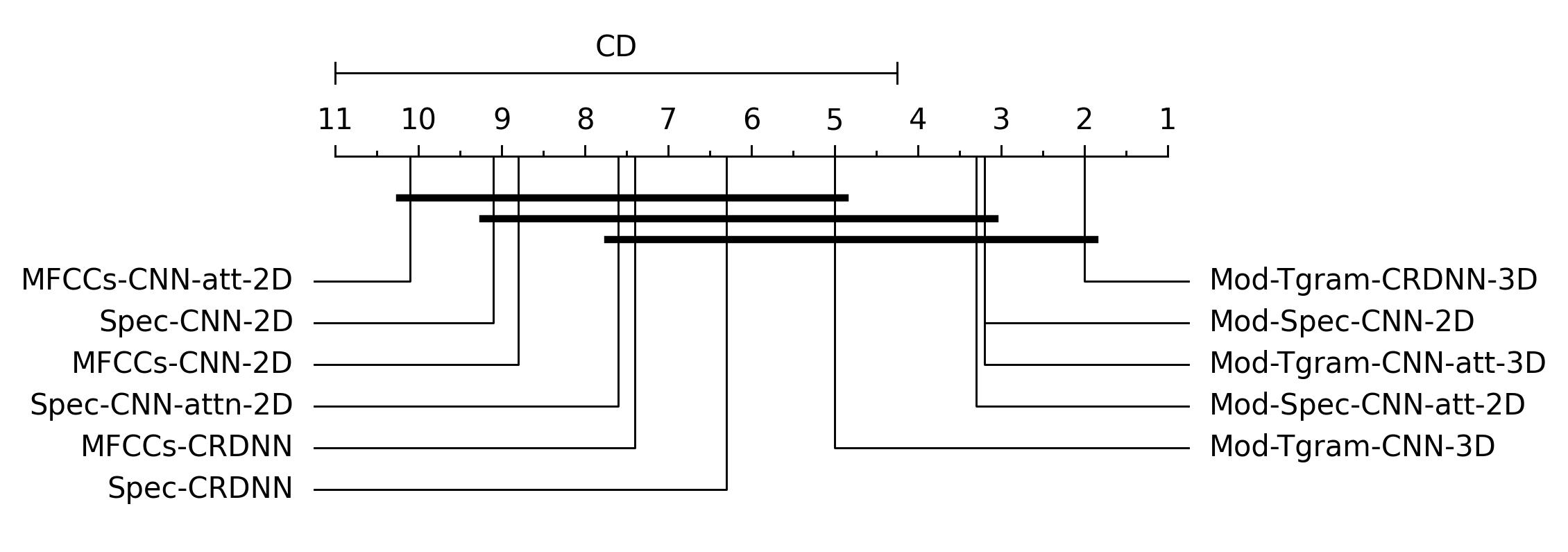}
}
\caption{Critical Difference (CD) plot illustrating the results of the Nemenyi post-hoc test following the Friedman test on the MAE for a) random-split and b) hive-independent scenario. The models and features combination are ranked by their average performance across the five folds. Features grouped by a thick horizontal bar exhibit no statistically significant difference in performance ($\alpha = 0.05$). Features with lower ranks (towards 1) demonstrate superior performance.}
\label{fig:cd_plots_mae}
\end{figure*}

Figures~\ref{fig:cd_plots_mae} (a)-(b) present the Critical Difference (CD) diagrams for the MAE metric under both the random-split and hive-independent evaluation setups, respectively. These diagrams summarize the post-hoc statistical comparisons: methods linked by a horizontal bar do not differ significantly, while those separated by more than the critical distance show statistically significant differences. As can be seen, in the random-split scenario, Mod-Tgram(tensorgram)-CRDNN-3D obtained the top average rank and outperformed models such as MFCCs-CNN-2D and Spec-CNN-att-2D, evidenced by the clear separation in the CD diagram. Conversely, in the hive-independent scenario, Mod-Tgram-CRDNN-3D  grouped distinctly from MFCCs-att-CNN-2D, highlighting its superior ability to generalize across hives. Several of the modulation‑based methods fall within the same statistical group, so the apparent ordering within that group should be interpreted as comparable performance rather than a strict ranking.

\subsection{Explainability analysis}
\begin{figure}%[H]
\centering
\subfloat[]{\label{fig:}
\centering
\includegraphics[width=0.8\linewidth]{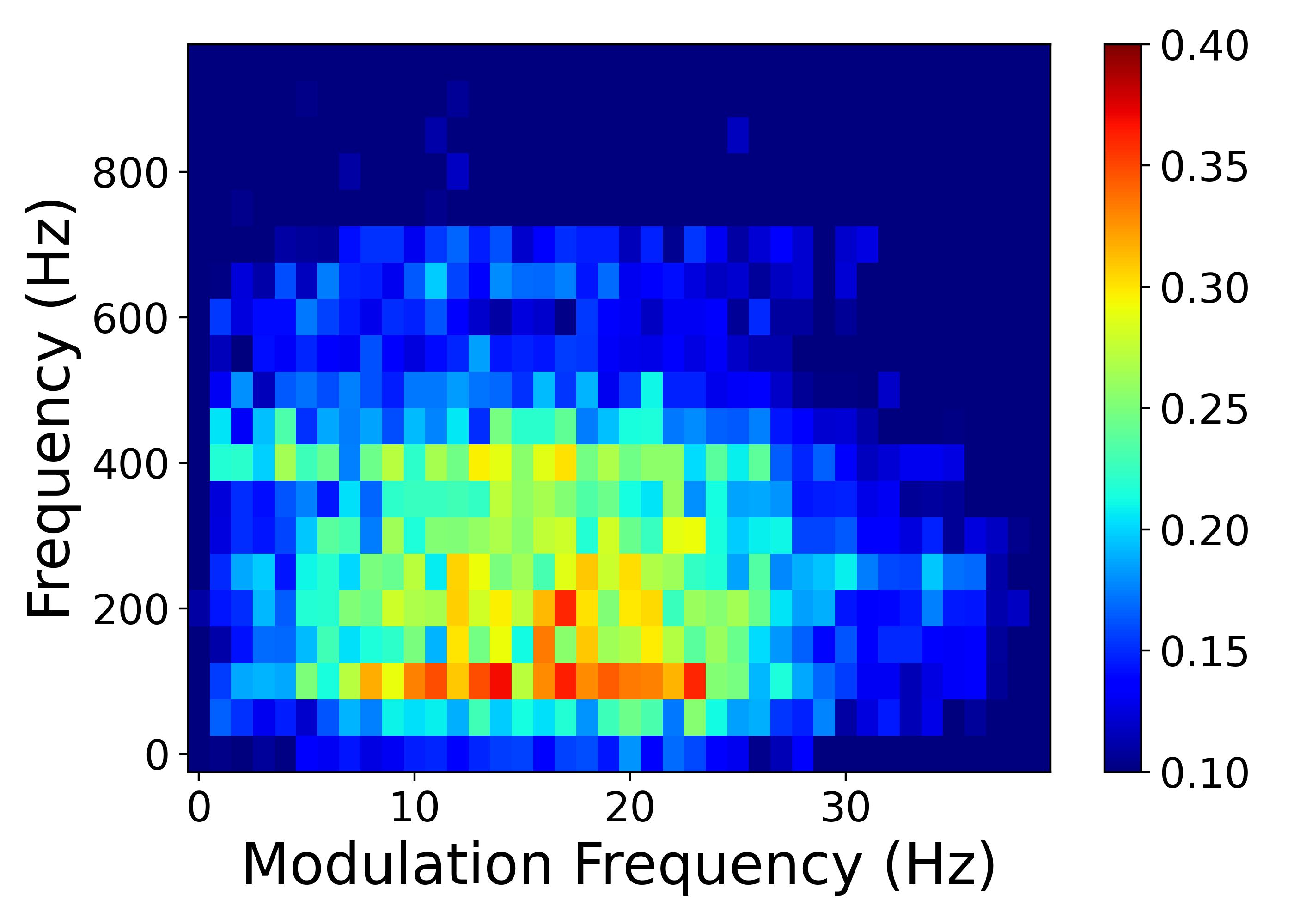 }
}

\subfloat[]{\label{fig:}
\centering
\includegraphics[width=0.8\linewidth]{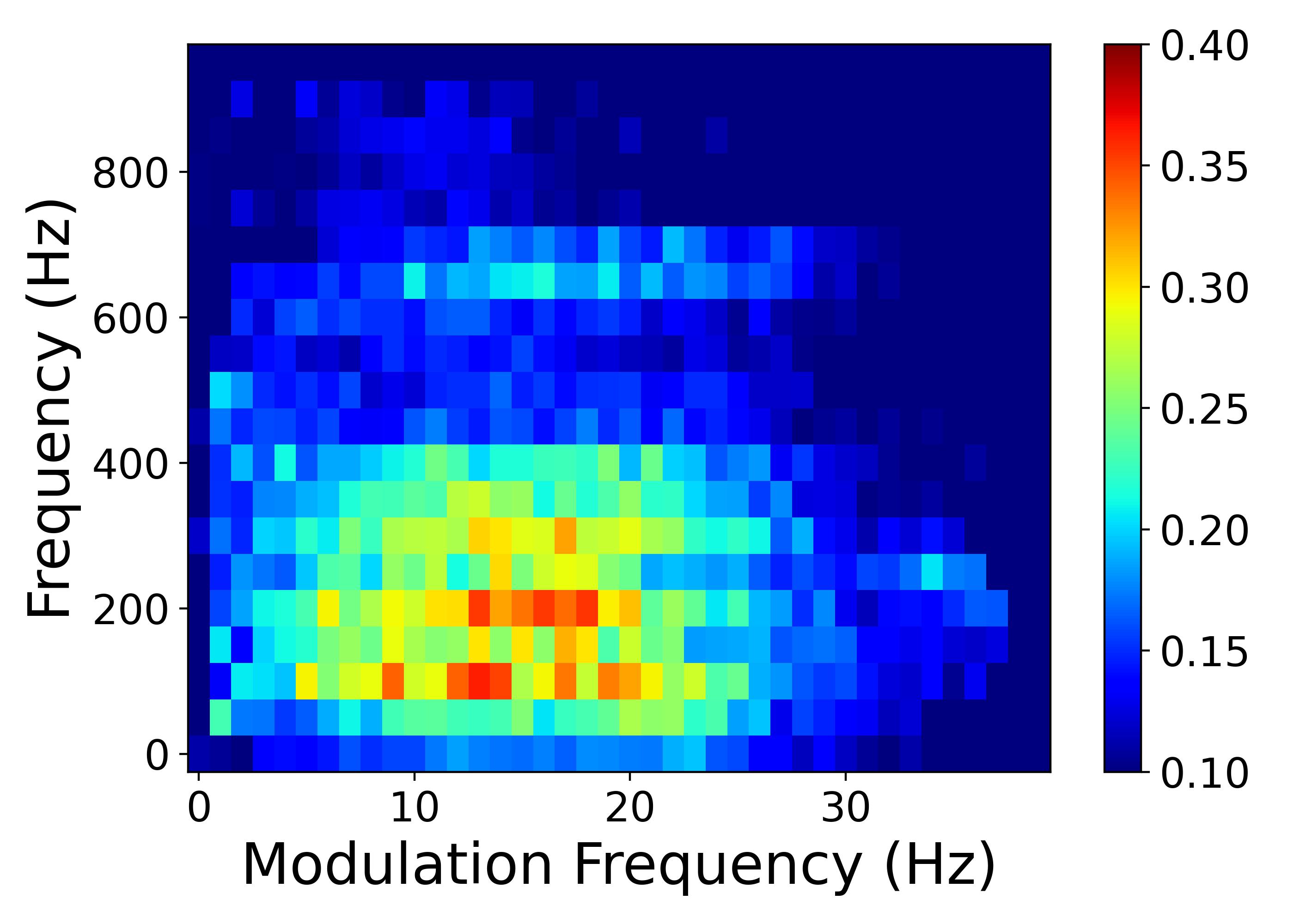}
}

\subfloat[]{\label{fig:}
\centering
\includegraphics[width=0.8\linewidth]{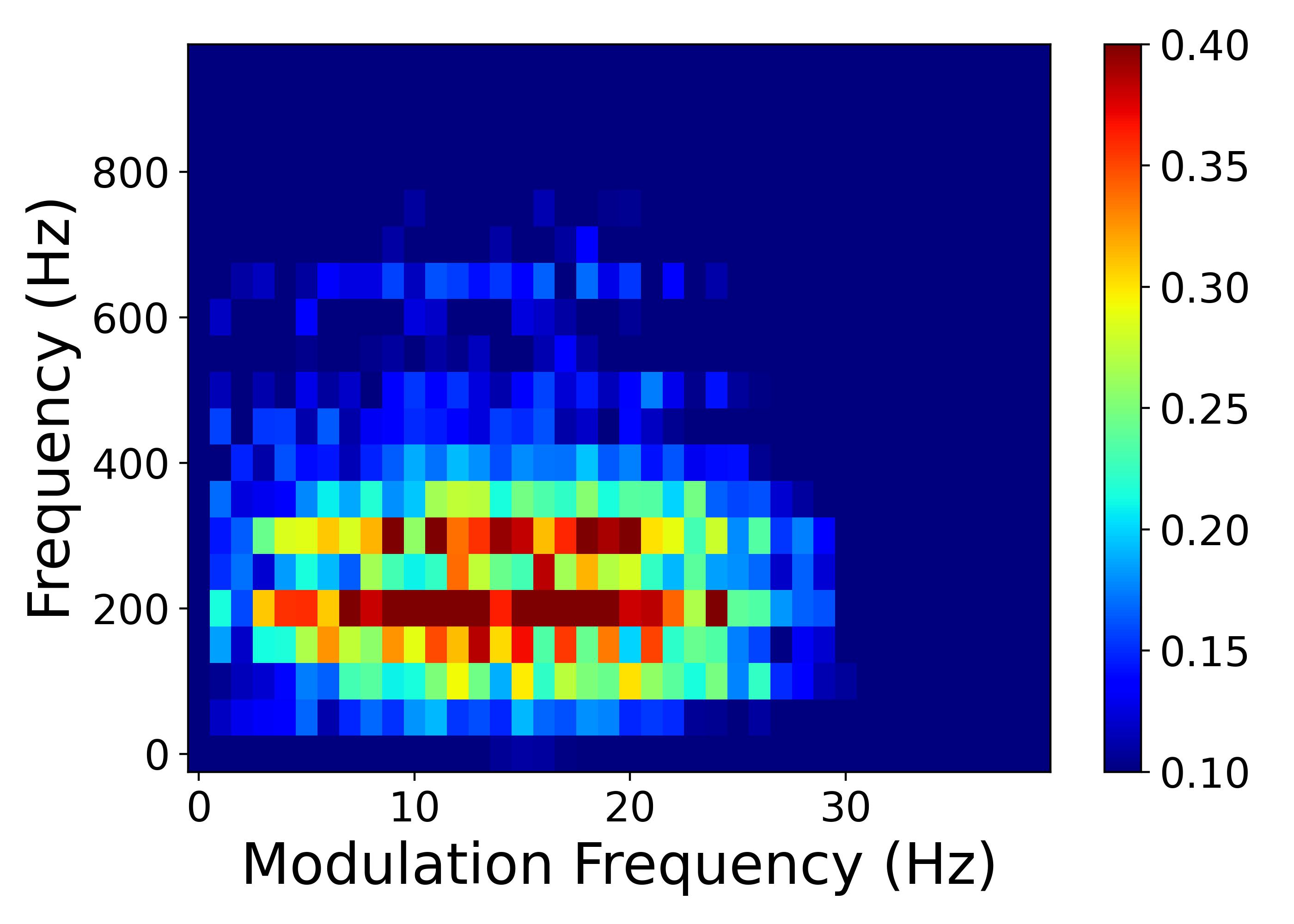}
}
\caption{Saliency maps of the trained 2D CNN model averaged over samples for each colony strength level: (a) low (1–10 fobs), (b) medium (11-20 fobs), and (c) high (21–30 fobs).} \vspace{-4mm}
\label{fig:saliency_cnn_2d}
\end{figure}

\begin{figure}
\centering
\subfloat[]{\label{fig:}
\centering
\includegraphics[width=0.8\linewidth]{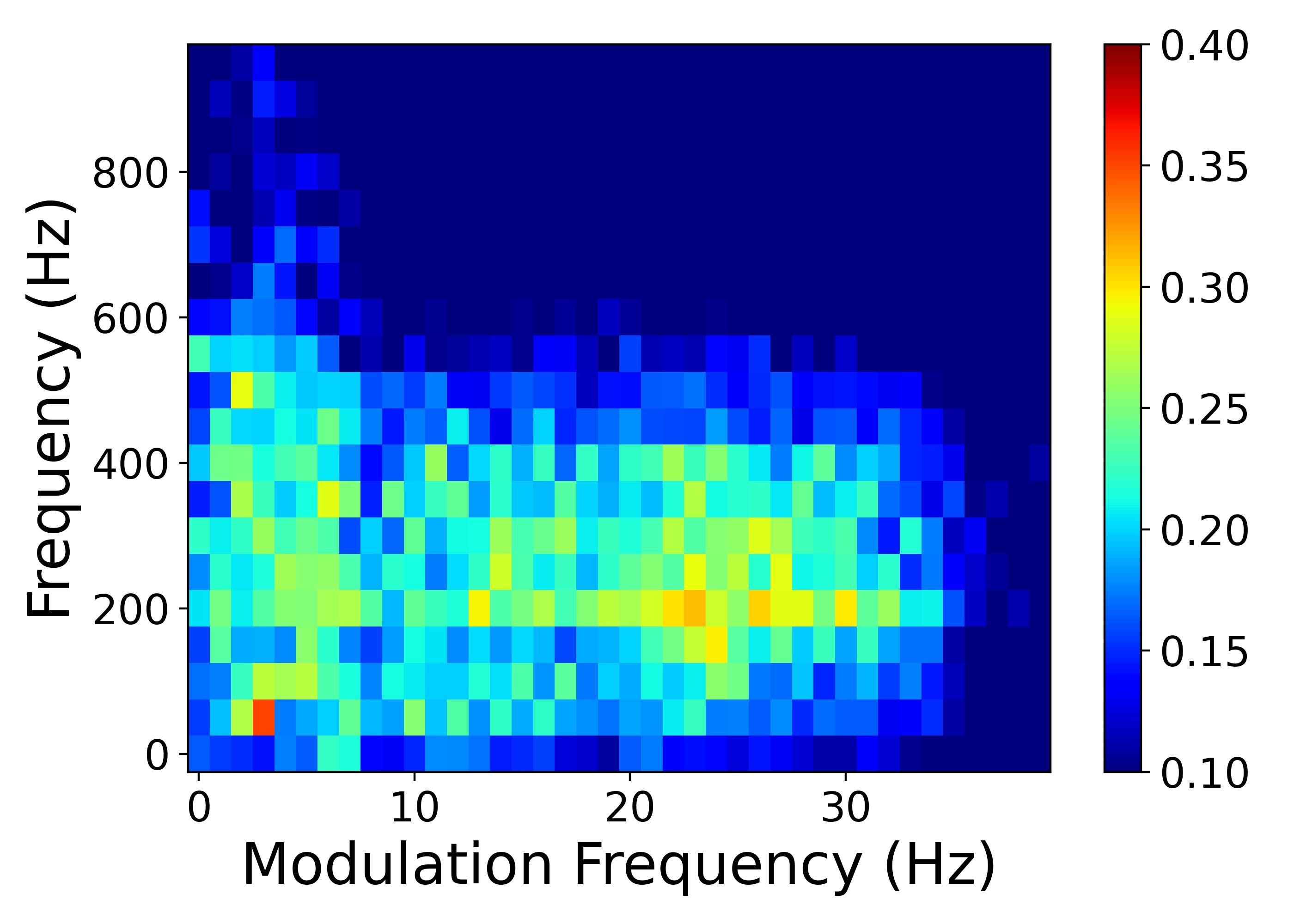 }
}

\subfloat[]{\label{fig:}
\centering
\includegraphics[width=0.8\linewidth]{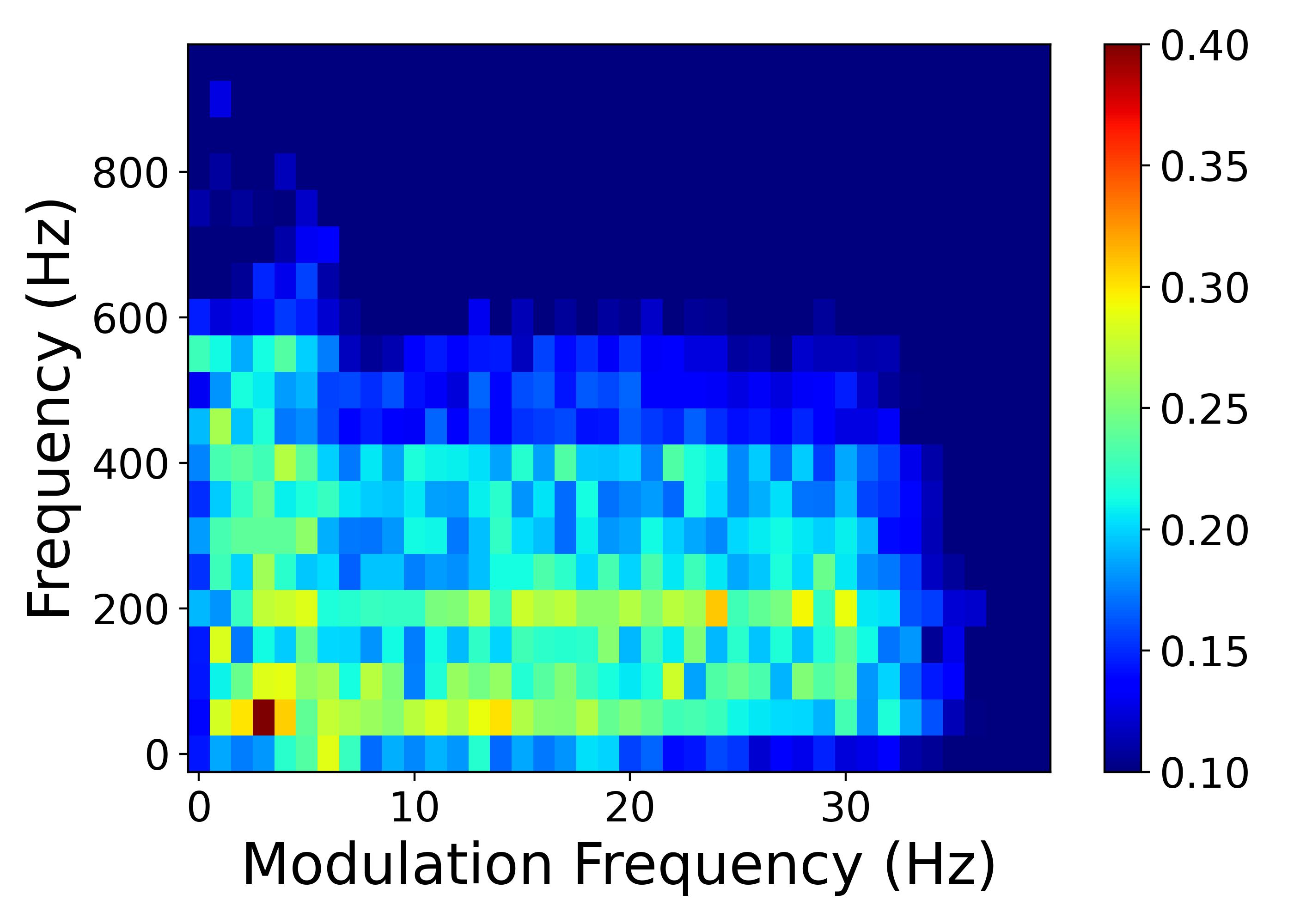}
}

\subfloat[]{\label{fig:}
\centering
\includegraphics[width=0.8\linewidth]{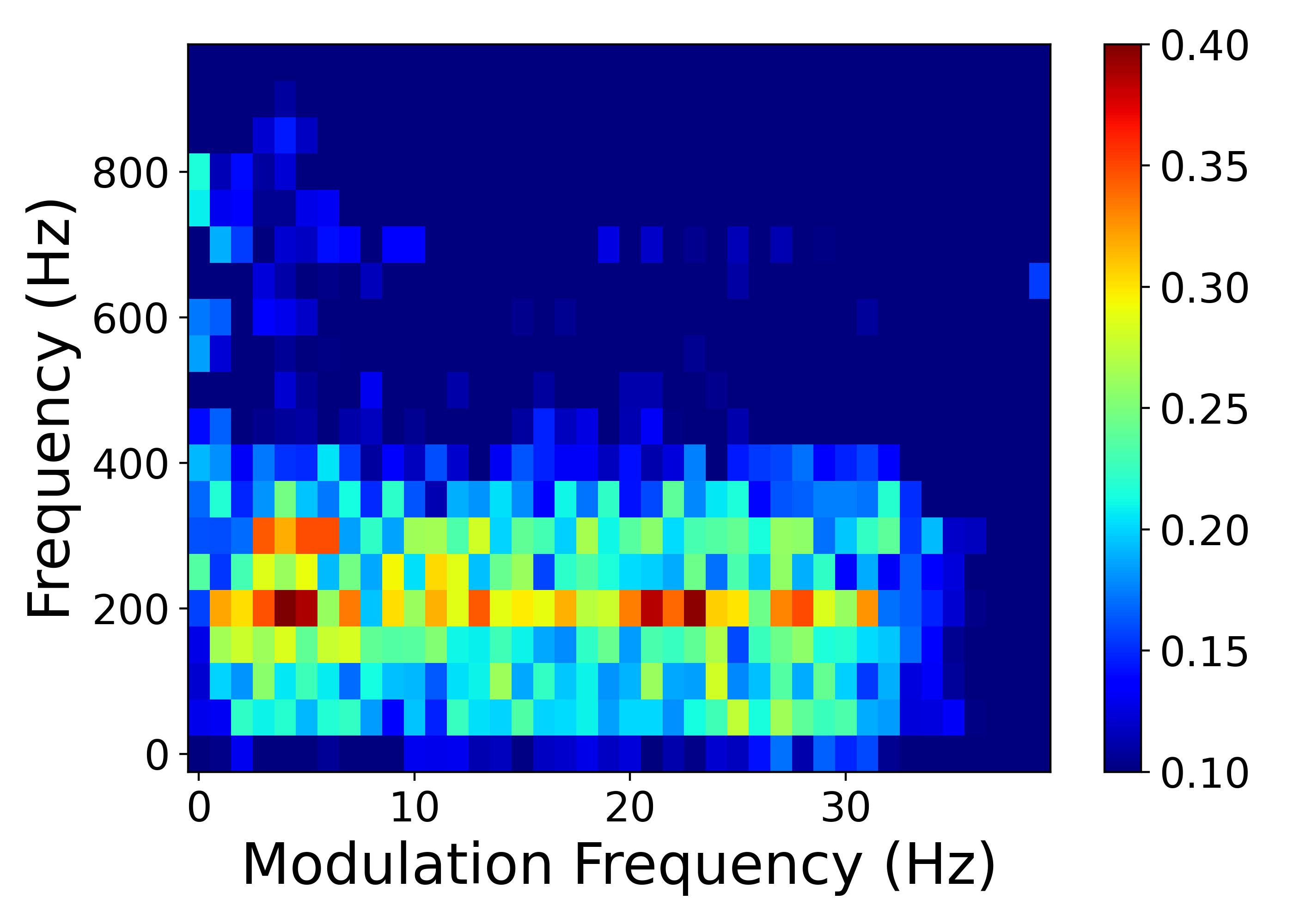}
}
\caption{Saliency maps of the trained 2D CNN model with spatial attention averaged over samples for each colony strength level: (a) low (1–10 fobs), (b) medium (11–20 fobs), and (c) high (21–30 fobs).}
\label{fig:saliency_cnn_2d_attention} \vspace{-4mm}
\end{figure}

To interpret the spatial and temporal relevance of the input modulation spectrograms, we generated saliency maps for the trained 2D CNN and attention-based models. Saliency maps highlight the input regions that most strongly influence the model’s predictions by computing the gradient of the output with respect to each feature. In our case, these maps reveal which frequency and modulation bands are most discriminative for different levels of bee activity. To reduce noise and capture consistent activation patterns, we averaged the saliency maps across multiple samples within each colony strength category (low, medium, and high). The resulting maps indicate that specific modulation frequency regions contribute more strongly as colony strength increases, suggesting that the model relies on structured acoustic patterns rather than isolated features. Figures~\ref{fig:saliency_cnn_2d} and~\ref{fig:saliency_cnn_2d_attention} show the saliency maps for the three strength levels for the 2D CNN and 2D CNN with spatial attention models, respectively.

The saliency maps for both the standard 2D CNN and the 2D CNN with spatial attention confirm that the models primarily rely on specific modulation-frequency patterns to predict bee activity levels, suggesting a reliance on structured acoustic patterns. For the standard 2D CNN (Figure~\ref{fig:saliency_cnn_2d}), increasing colony strength correlates with a progressive expansion and upward shift of the relevant frequency region, suggesting that higher harmonics or broader spectral features become relevant. In contrast, the 2D CNN with spatial attention (Figure~\ref{fig:saliency_cnn_2d_attention}) maintains a consistently narrower, more horizontal high-saliency band, primarily focused around 200\,Hz, which expands across the modulation frequency axis as activity increases. This suggests the attention mechanism is effective at isolating the single most discriminative fundamental frequency band and focusing on its temporal variation (modulation), whereas the standard CNN integrates a broader range of adjacent frequencies. Notably, both models show a peak relevance in the 10--25\,Hz modulation frequency range across the medium and high strength categories, indicating that these specific rates of temporal fluctuation are crucial discriminative features.

Next, for explainability of the 3D CNN model, we used the gradient-weighted class activation mapping (Grad-CAM) method \cite{selvaraju2017grad}. Grad-CAM leverages intermediate feature activations to generate interpretable maps across spatial and temporal dimensions. By averaging these maps across samples, we identified the regions most relevant to the model’s predictions, revealing how specific frequency–modulation patterns evolve with colony strength. This approach allowed us to visualize and compare differences in the 2D and 3D architectures.

\begin{figure*}
    \centering
    
    %---------------- ROW 1: t1 ----------------%
    \begin{subfigure}{0.32\textwidth}
        \centering
        \includegraphics[width=0.9\linewidth]{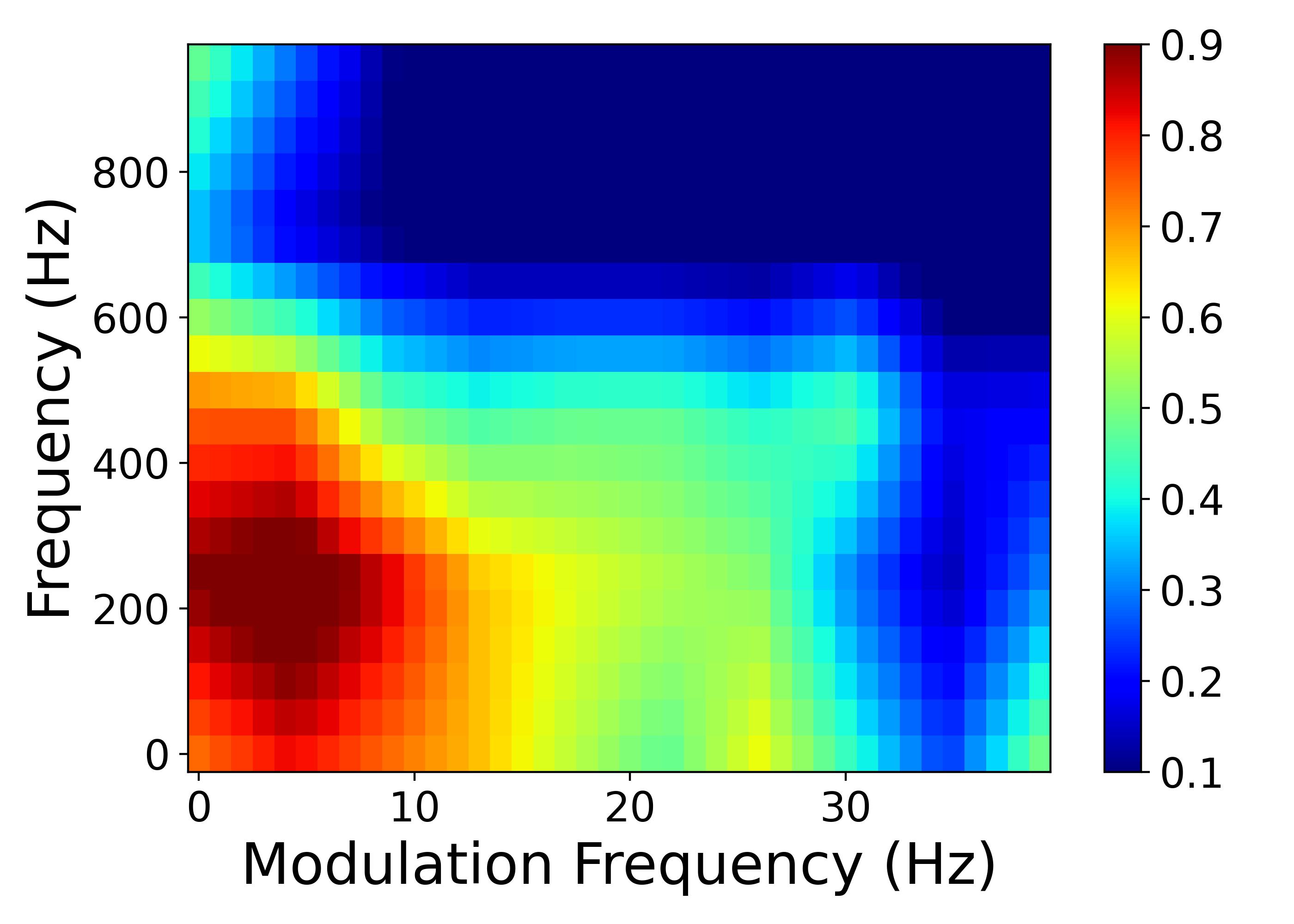}
        \caption{Low (1--10 fobs).}
    \end{subfigure}%
    \hfill
    \begin{subfigure}{0.32\textwidth}
        \centering
        \includegraphics[width=0.9\linewidth]{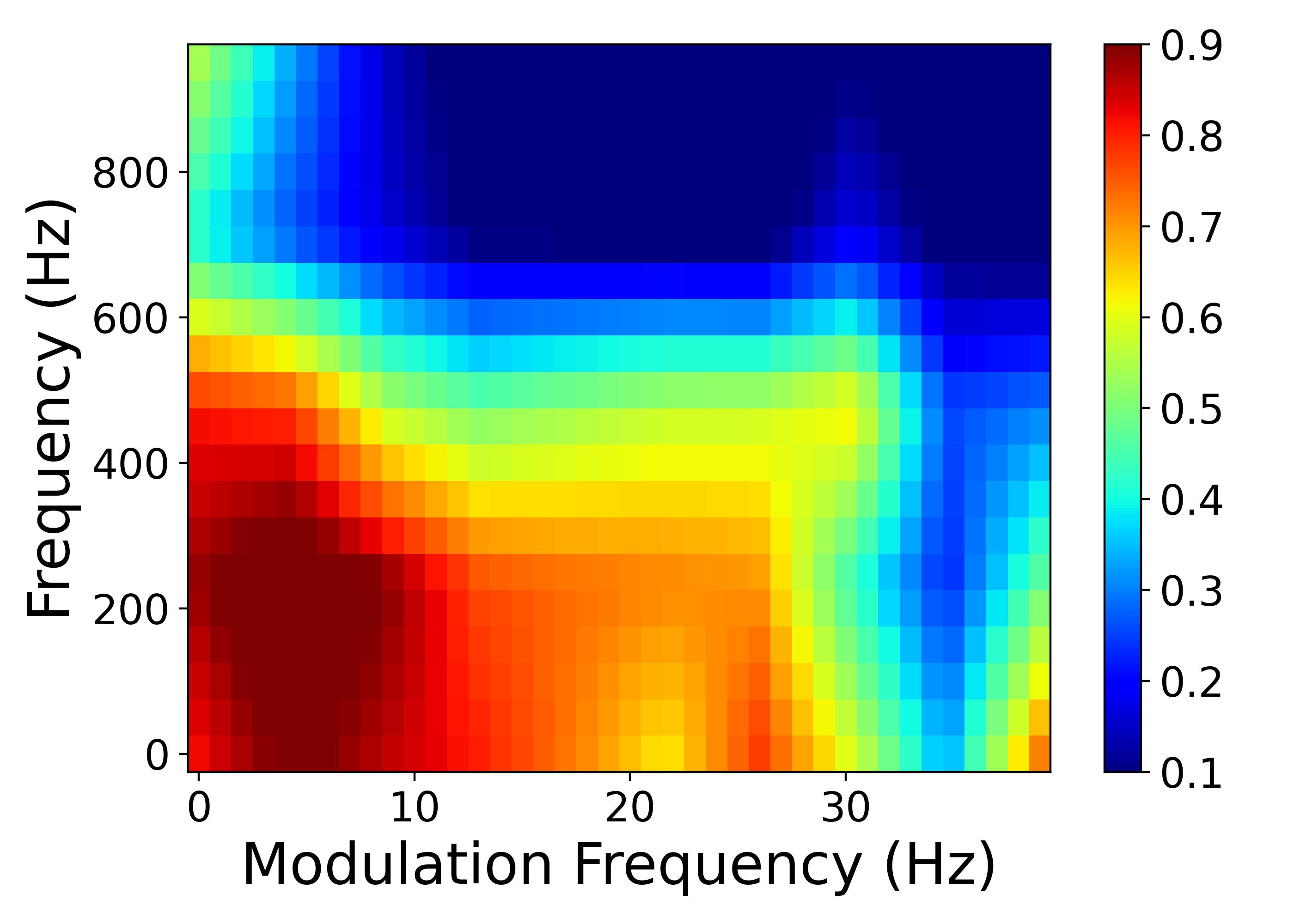}
        \caption{Medium (11--20 fobs).}
    \end{subfigure}%
    \hfill
    \begin{subfigure}{0.32\textwidth}
        \centering
        \includegraphics[width=0.9\linewidth]{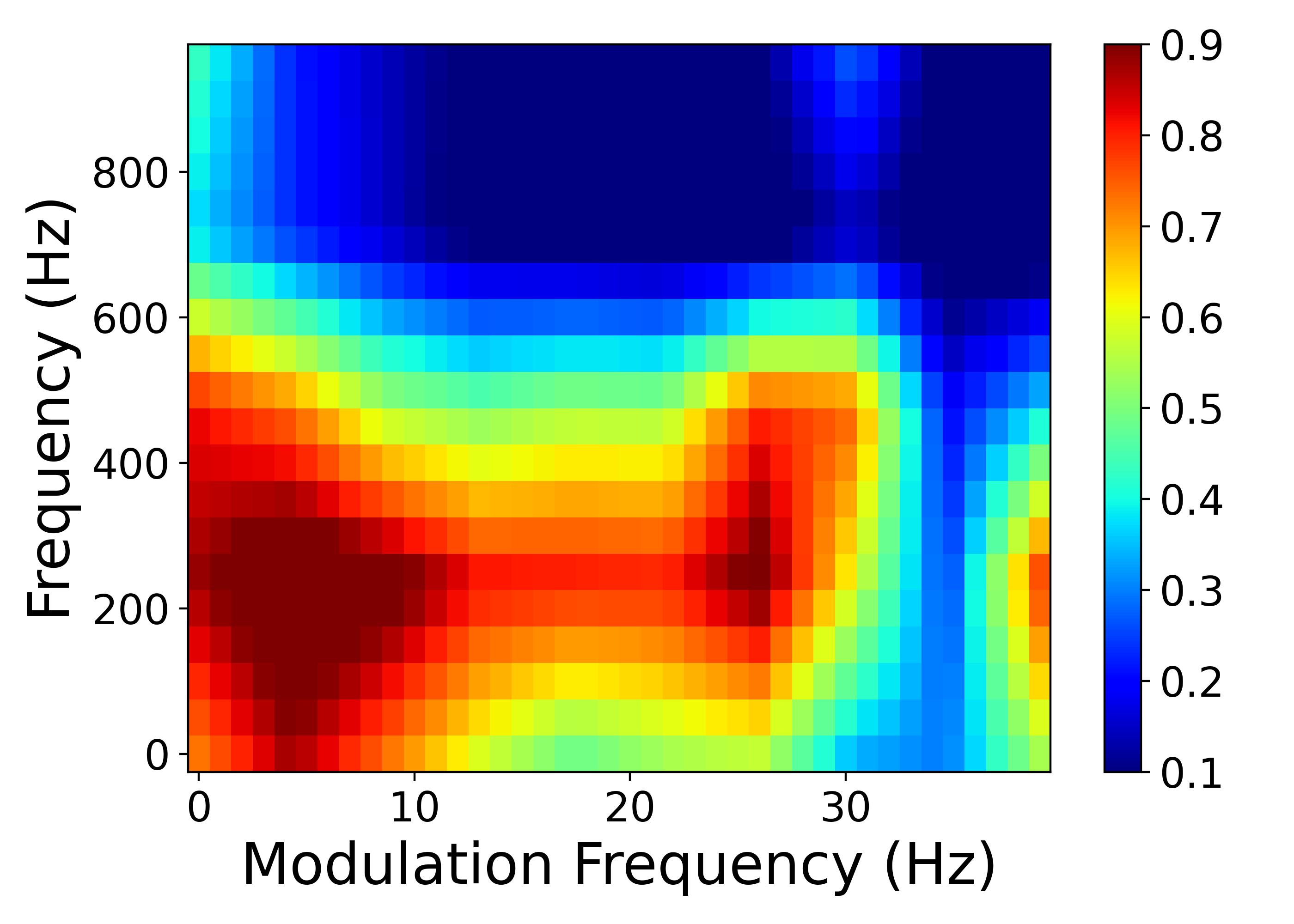}
        \caption{High (21--30 fobs).}
    \end{subfigure}

    \vspace{8pt}
    \textbf{Time frame: $t_1$} \\[12pt]

    %---------------- ROW 2: tmid ----------------%
    \begin{subfigure}{0.32\textwidth}
        \centering
        \includegraphics[width=0.9\linewidth]{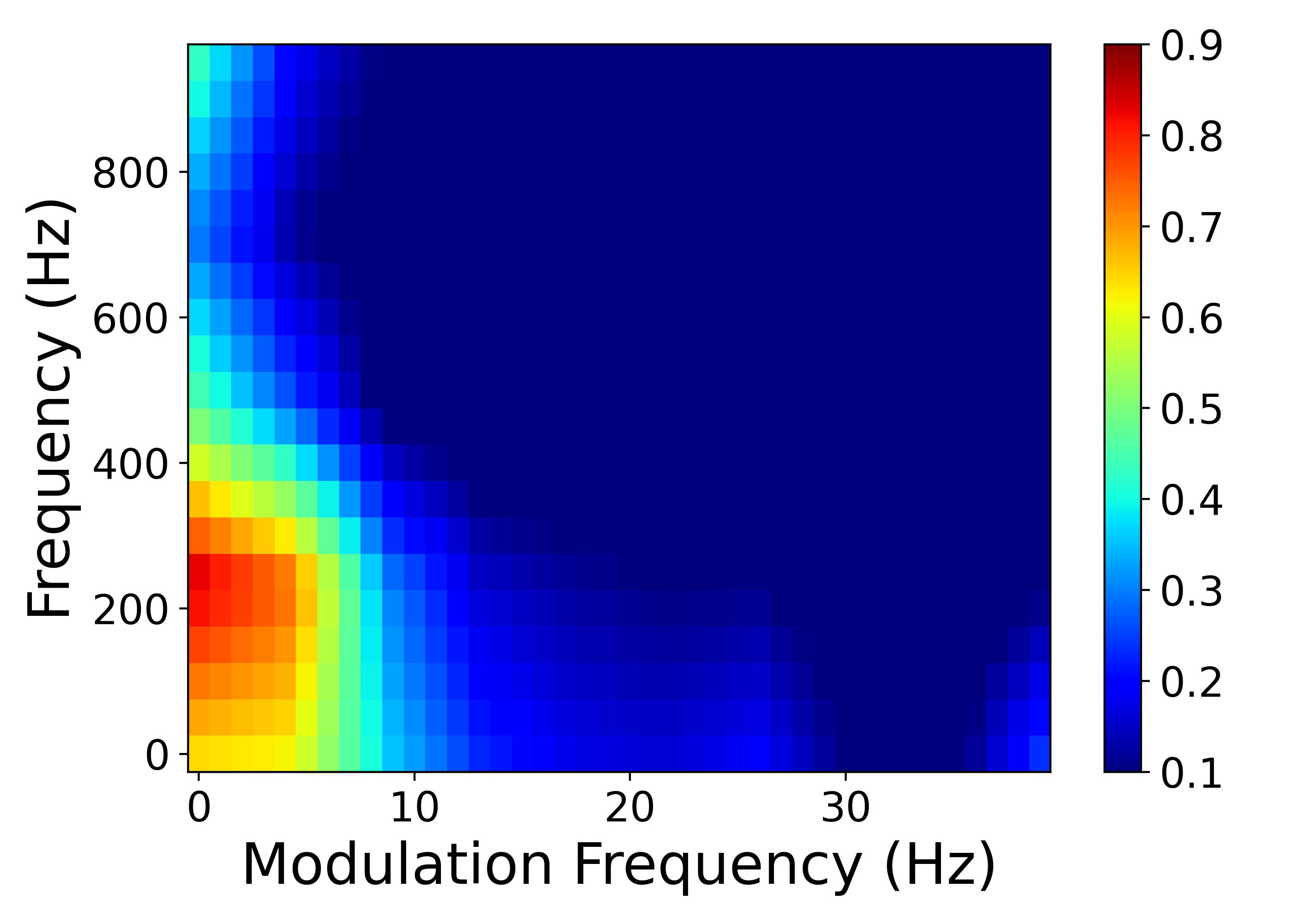}
        \caption{Low.}
    \end{subfigure}%
    \hfill
    \begin{subfigure}{0.32\textwidth}
        \centering
        \includegraphics[width=0.9\linewidth]{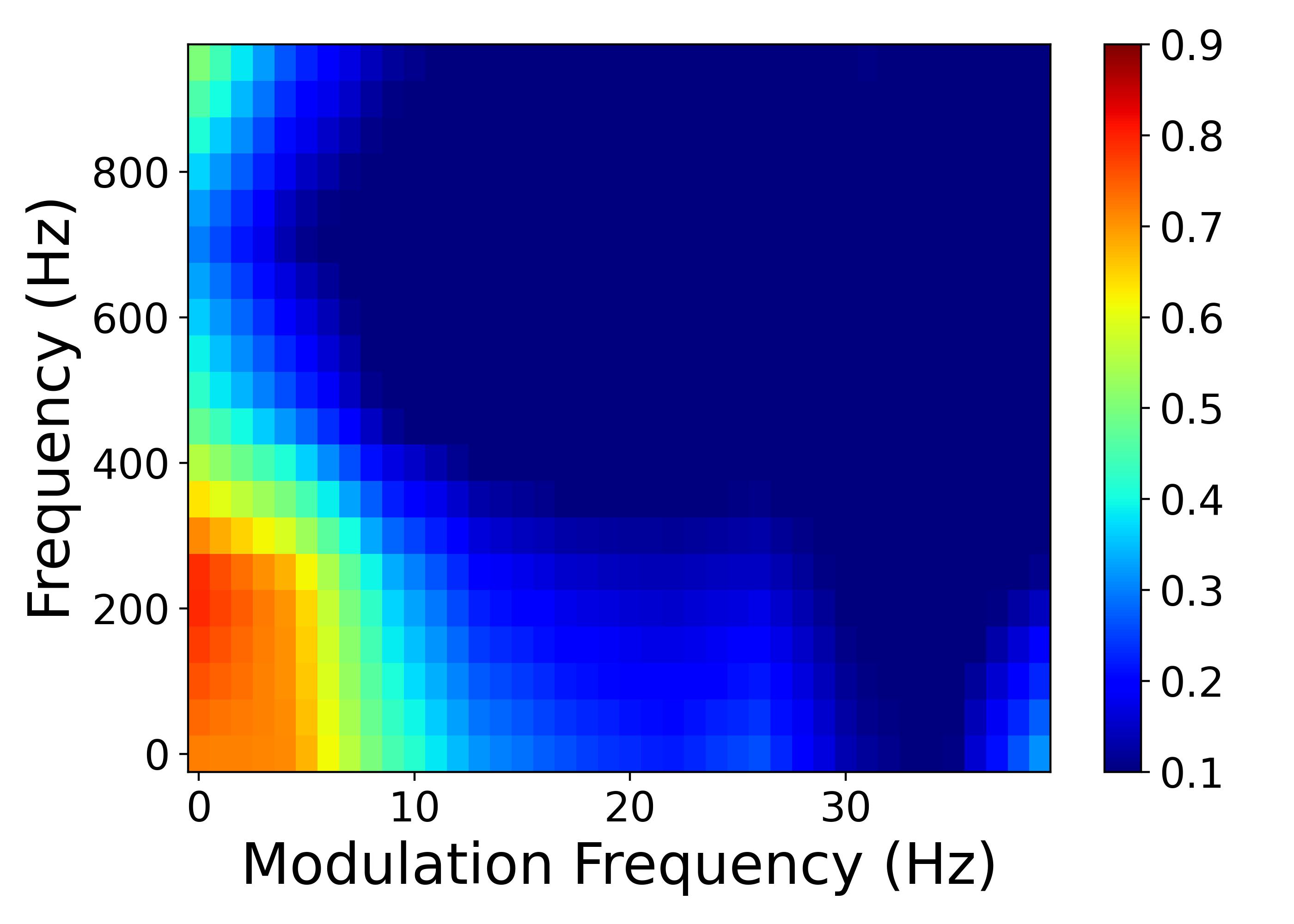}
        \caption{Medium.}
    \end{subfigure}%
    \hfill
    \begin{subfigure}{0.32\textwidth}
        \centering
        \includegraphics[width=0.9\linewidth]{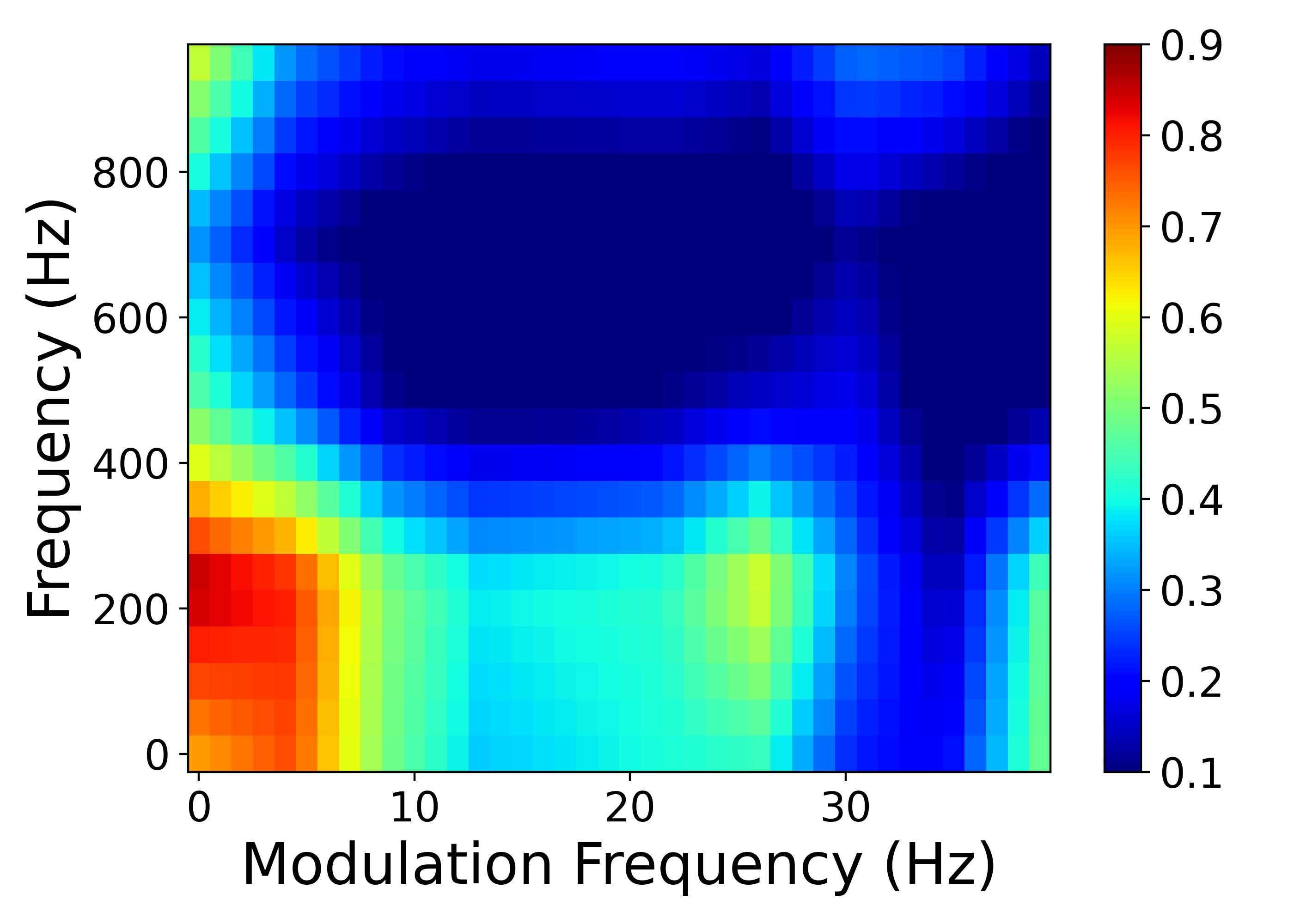}
        \caption{High.}
    \end{subfigure}

    \vspace{8pt}
    \textbf{Time frame: $t_{\mathrm{mid}}$} \\[12pt]

    %---------------- ROW 3: tmax ----------------%
    \begin{subfigure}{0.32\textwidth}
        \centering
        \includegraphics[width=0.9\linewidth]{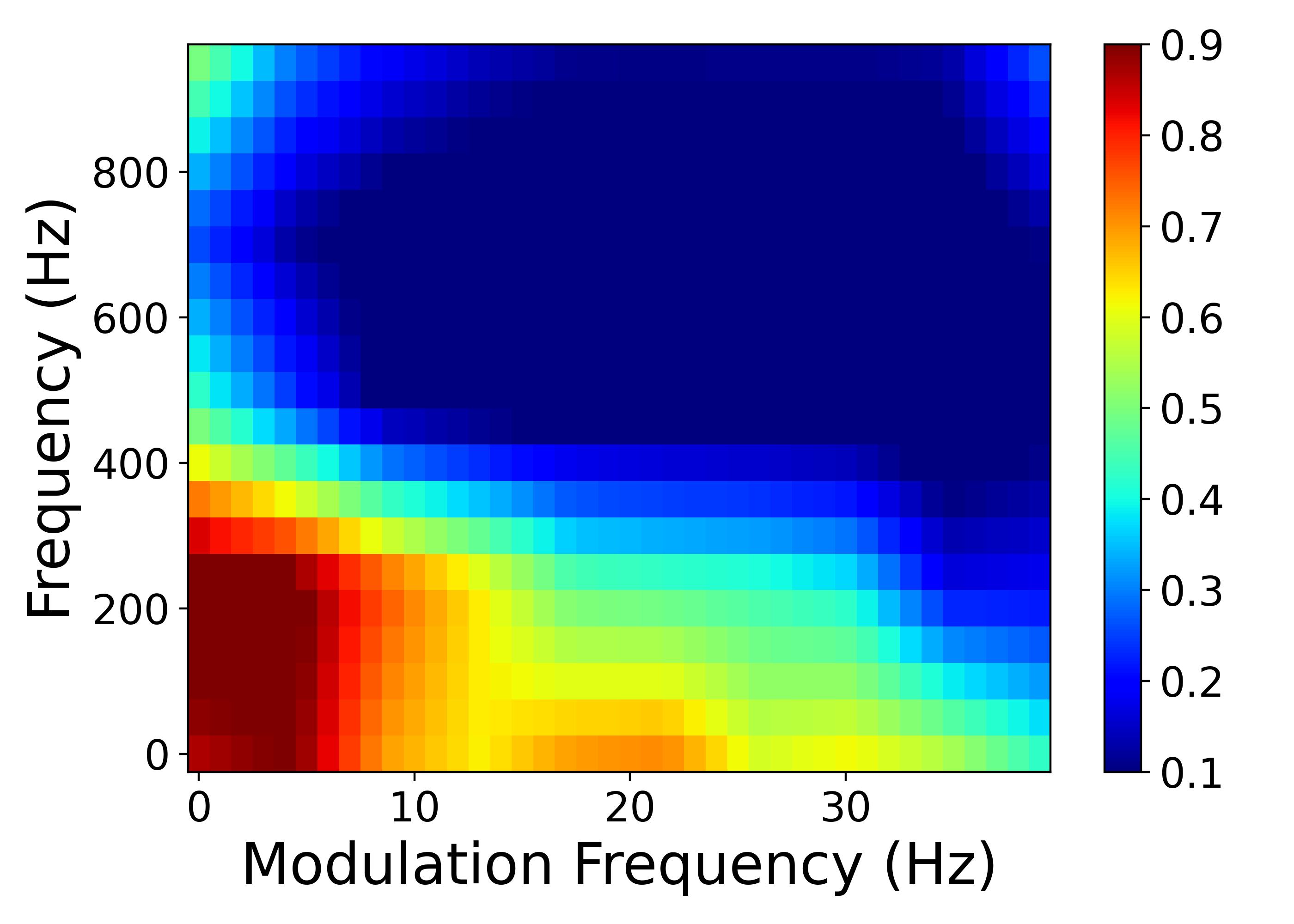}
        \caption{Low.}
    \end{subfigure}%
    \hfill
    \begin{subfigure}{0.32\textwidth}
        \centering
        \includegraphics[width=0.9\linewidth]{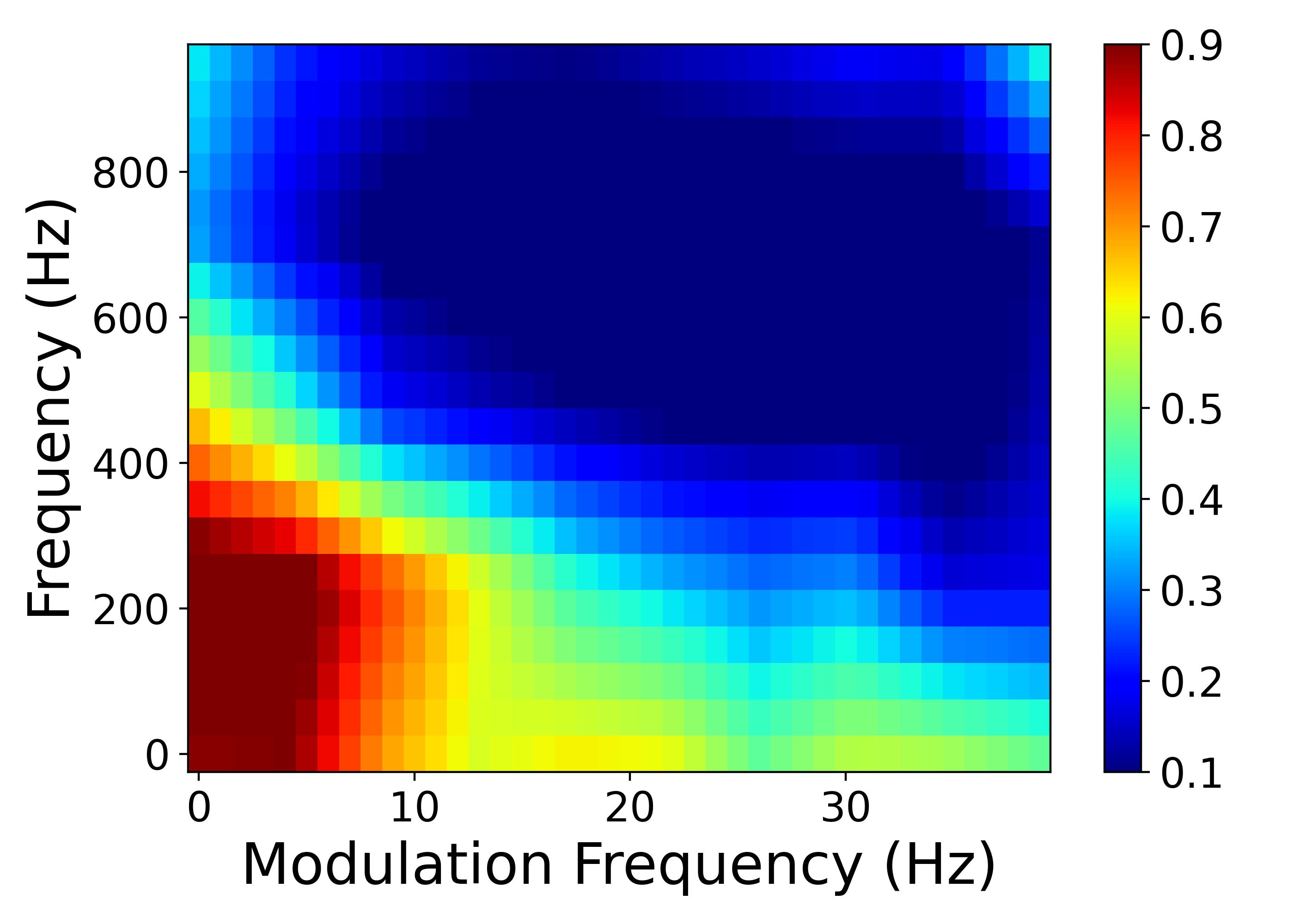}
        \caption{Medium.}
    \end{subfigure}%
    \hfill
    \begin{subfigure}{0.32\textwidth}
        \centering
        \includegraphics[width=0.9\linewidth]{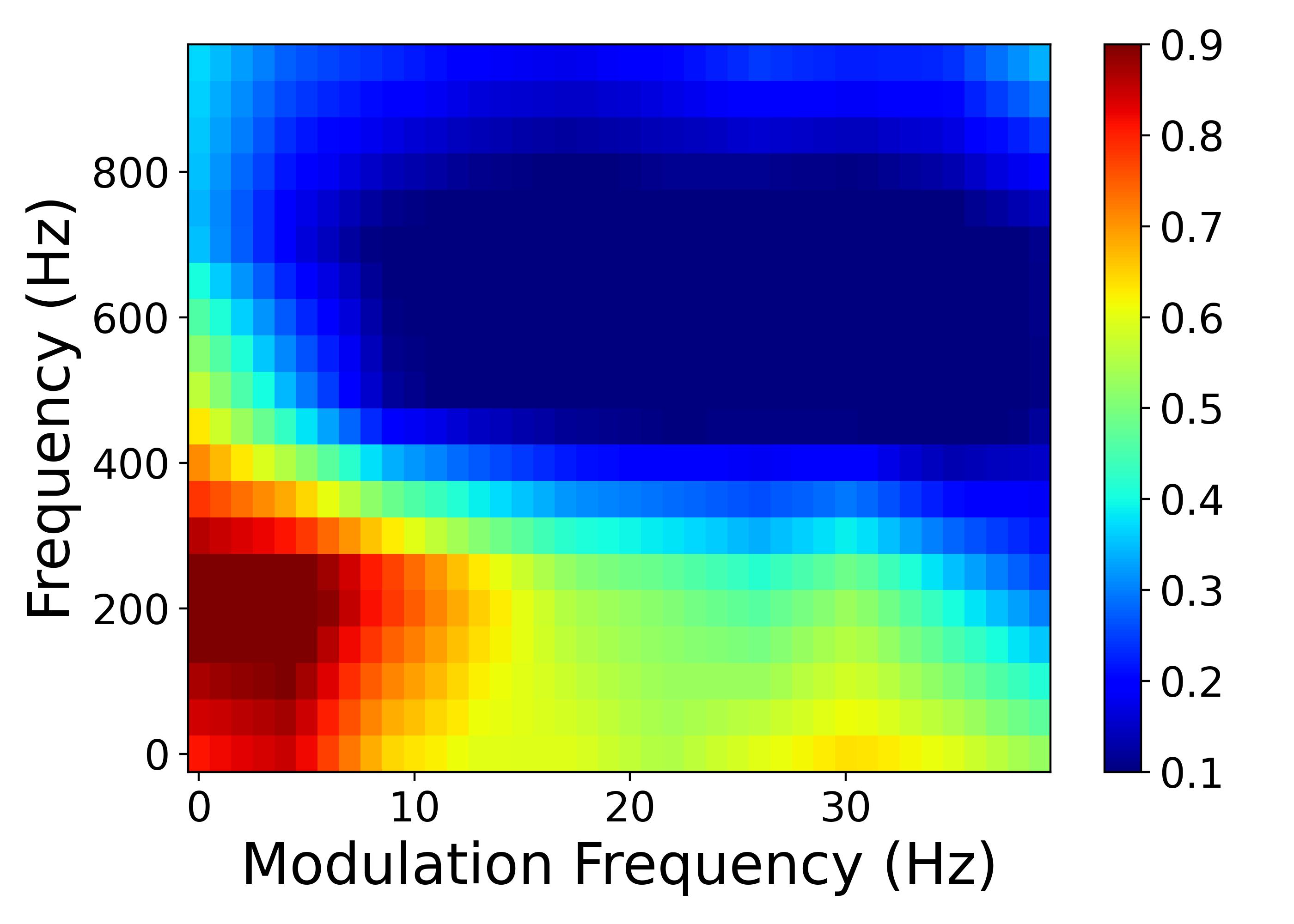}
        \caption{High.}
    \end{subfigure}

    \vspace{8pt}
    \textbf{Time frame: $t_{\mathrm{max}}$}

    \caption{
        Grad-CAM visualizations for representative 3D test samples.  
        Each row corresponds to a different time frame ($t_1$, $t_{\mathrm{mid}}$, $t_{\mathrm{max}}$),  
        and columns correspond to colony strength levels (low, medium, high).  
        Heatmaps show how the model’s focus shifts across the modulation-frequency space over time.
    }
    \label{fig:3d_gradcam} \vspace{-4mm}
\end{figure*}

Rather than focusing solely on a single aggregated explanation, we examined several representative frames including the first, middle, and last frames to capture the model’s response to temporal variations in the modulation spectrogram. We found that high-activity samples exhibit more consistent and spatially distributed activation across frames, while low-activity samples display more localized and intermittent patterns. These observations suggest that the 3D models leverage both spectral structure and temporal evolution to estimate colony strength levels. Figure~\ref{fig:3d_gradcam} provides a detailed visualization of these temporal dynamics. For each strength level (left column=low, middle=medium, right=high), the figure shows the three Grad-CAM heatmaps extracted from three different time frames (top row=first, middle=middle, bottom=last frame).

As can be seen, for low strength colonies (1–10 fobs), the model's focus is generally localized to low frequencies (below 400\,Hz) and very low modulation frequencies (below 10\,Hz), becoming notably intermittent and contracting severely at the middle frame ($t_{\mathrm{mid}}$). In contrast, high strength colony (21–30 fobs) samples show a consistent and broadly distributed activation across all frames, leveraging low-to-mid frequencies (up to 550\,Hz) and spanning the entire range of modulation frequencies (up to 35\,Hz), which suggests the model uses a stable, widespread feature set. The medium strength class (11–20 fobs) exhibits a dynamic pattern, shifting from a broad focus in $t_1$ to a localized one in $t_{\mathrm{mid}}$, and then expanding again in $t_{\mathrm{max}}$. These observations confirm that the 3D architecture effectively leverages both spectral structure (Frequency) and its temporal evolution (Modulation Frequency over time) to differentiate colony strength levels.

\subsection{Deep learning vs. Conventional classifiers}
As a last exploration, we wish to compare the results obtained with the deep learning methods described herein with those of our earlier work that relied on a conventional random forest (RF) classifier \cite{abdollahi2025audio}. Table \ref{tab:rf_dl_comparison} summarizes the results obtained with the best RF model and ours. As can be seen, the proposed method improves accuracy across both figures-of-merit in the more stringent hive-independent scenario. Particularly, the proposed CRDNN-3D model is able to take advantage of the modulation tensorgram 3D input to better characterize the temporal changes of the bee buzzing sounds, something that the RF classifier with handcrafted features is not able to achieve.

\begin{table}%[h]
    \centering
    \caption{Performance comparison of best-performing random forest (RF) classifier and best-performing deep learning (DL) method presented herein in the hive-independent scenario.}
    \label{tab:rf_dl_comparison}
    \begin{tabular}{l|c|c}
        \toprule
        \textbf{Metric} & \textbf{RF Best (MSAB)} & \textbf{DL Best (CRDNN-3D)} \\
        \midrule
        MAE & 3.41 $\pm$ 1.03 & 3.31 $\pm$ 1.36 \\
        \rr & 0.74 $\pm$  0.15 & 0.78 $\pm$ 0.17 \\
        \bottomrule
    \end{tabular}
\end{table}

\subsection{Computational Complexity}\label{Computation}

The computational complexity results, summarized in Table~\ref{tab:complexity}, were obtained on a MacBook Pro equipped with an Apple M1 chip and 16 GB of unified memory. Model sizes were extracted directly from the saved .h5 files, while inference times correspond to the average runtime over multiple forward passes per model.

Overall, 2D CNN-based models offer the most efficient inference, with CNN-2D achieving the lowest runtime (0.29 ms) and a compact footprint (6.1 MB). The addition of spatial attention slightly increases computational cost (0.41 ms), while remaining lightweight and suitable for real-time IoT deployment.
3D architectures are more computationally demanding due to volumetric feature extraction. CNN-3D and CNN-att-3D exhibit higher inference times (3.93 ms and 10.70 ms, respectively) and larger model sizes, reflecting the added cost of 3D convolutions and attention mechanisms.

In contrast, CRDNN models provide a favorable balance between efficiency and representational power. Although CRDNN-3D has a higher inference time than 2D CNNs (4.71 ms), it remains significantly more efficient than CNN-att-3D while achieving the best overall predictive performance. This suggests that the recurrent component effectively captures temporal dependencies in the data, leading to improved accuracy without an excessive increase in model complexity. Notably, CRDNN-3D also maintains a relatively moderate model size (4.5 MB), making it a strong candidate for practical deployment. Models with size below ~10 MB and inference under ~5 ms on a consumer CPU (CRDNN‑2D, CNN‑2D, CNN‑att‑2D, CRDNN‑3D) are plausible candidates for on‑hive edge deployment; the heavier 3D CNNs (CNN‑3D, CNN‑att‑3D) are more naturally deployed cloud‑side with the IoT sensor uploading raw audio.

\begin{table}
\centering
\caption{Computational complexity of the evaluated deep learning models.}
\label{tab:complexity}
%\resizebox{0.8\textwidth}{!}{
\begin{tabular}{lcccc}
\toprule
\textbf{Model} & \textbf{\# Parameters} & \textbf{\makecell{Model Size\\ (MB)}}  & \textbf{\makecell{Inference Time\\ (ms)}} \\
\midrule
CNN-2D                      & 503,553 &  6.1 & 0.29  \\
CNN-att-2D                 &  503,850 &  6.2 & 0.41  \\
CRDNN-2D                   & 179,137 & 2.3 &  0.62 \\
CNN-3D                     & 1,507,649 & 18.2 &  3.93 \\
CNN-att-3D                 & 1,919,310 & 23.2 &  10.70 \\
CRDNN-3D                   & 364,033 & 4.5 & 4.71\\
\bottomrule
\end{tabular}
%}
\end{table}

\subsection{Limitations and Future Directions}\label{limitations}

One limitation of the present study is that colony strength may correlate with several external and hive-specific factors beyond the intrinsic acoustic activity of the bees. Variables such as seasonality, weather conditions, time of day, hive configuration, and colony-specific characteristics may influence the recorded acoustic patterns and therefore contribute to the predictive behavior of the models. Although a hive-independent evaluation protocol was adopted to reduce the possibility of learning hive-specific acoustic signatures, temporal and seasonal confounding effects may still remain. For example, colony population naturally varies across seasons alongside changes in brood activity, foraging behavior, and environmental conditions, which may indirectly influence the learned representations. Similarly, differences in hive configuration, such as brood chamber-only colonies versus colonies with additional supers, may alter the acoustic environment.
Finally, although the differential effect of spectral subtraction on baseline vs. modulation‑based features suggests improved noise robustness for the proposed representation, a direct evaluation conditioned on labelled environmental noise events (rain, wind, speech, vehicles) could not be performed on the UrBAN dataset, which does not provide per‑segment noise annotations.

Future work will therefore focus on improving the robustness and interpretability of the proposed framework under more diverse and controlled conditions. This includes season-stratified evaluation protocols, the incorporation of explicit environmental covariates such as temperature and humidity, normalization strategies accounting for circadian acoustic variations, and domain adaptation techniques to better separate colony strength–related acoustic signatures from environmental or seasonal effects.

\section{Conclusion}\label{conclusion}
In this study, we proposed the use of modulation spectrograms and tensorgrams, derived from beehive acoustic recordings, as input to deep learning neural network architectures for improved honey bee colony strength prediction. With a recent open dataset of raw audio beehive recordings, we compared the performance of two- and three-dimensional CNNs and their attention-enhanced variants, as well as convolutional recurrent CNNs. Results show that the proposed modulation tensorgrams and modulation spectrograms provide richer information for model learning, leading to improved and more generalizable prediction accuracy compared to conventional audio features. These results suggest that modulation spectral signal processing is a viable tool for automated in-the-wild beehive monitoring. 

\section*{Acknowledgments}
The authors acknowledge funding from NSERC via their Alliance program (ALLRP 548872-19), as well as Nectar Technologies Inc and Evan Henry for the support with the collection of the UrBAN dataset.

%\bibliographystyle{IEEEtran}
%\bibliography{sample}

\end{document}